\documentclass{article}
\usepackage{amsmath}
\usepackage{latexsym}
\usepackage{mathrsfs}
\usepackage{geometry}
\title{Rescuing the Born Rule for Quantum Cosmology}
\author{Joshua H. Cooperman\\ \emph{Department of Physics, University of California, Davis, CA 95616}}

\begin{document}
\maketitle

\begin{abstract}
Page has recently argued that the Born rule does not suffice for computing all probabilities in quantum cosmology. He further asserts that the Born rule's failure gives rise to the cosmological measure problem \cite{DNP}. Here I contend that Page's result stems from his use of an overly restrictive definition of the Born rule. In particular, I demonstrate that all of the probabilities he wishes to compute follow from the Born rule when generalized measurements are permitted. I also register two comments on Page's theoretical setting, relating respectively to Hilbert space dimensionality and permutation symmetry. These considerations lead me to conclude that the claimed insufficiency of the Born rule is by no means specific to the cosmological context.
\end{abstract}

\section{Setting the Scene}
Envision the following scenario. We live in a very large universe (or multiverse depending on your semantic persuasion). In fact, our universe is so vast that presumably there exist multiple copies of exactly identical systems at different spacetime locations.\footnote{Certain cosmological scenarios seem to suggest situations of this sort. See, for instance, reference \cite{SW} for a discussion of such scenarios and of the cosmological measure problem. While I hope that a more complete theory will militate against such scenarios, we cannot yet rule them out.} As local observers in the universe, we know neither which copy of ourselves we are nor how many copies of ourselves there are. As intrepid physicists in the universe, we wish to study its quantum mechanics. Page contends that we must confront a singular difficulty: the Born rule of quantum theory will fail for the computation of certain probabilities that we would desire to compute \cite{DNP}.\footnote{The remainder of this section comprises a review of the argument given in the fourth section of reference \cite{DNP}.}

Page defines the Born rule as the statement that all probabilities in quantum theory are computed as expectation values of a complete set of orthogonal projection operators. Symbolically, for a system in the state $|\psi\rangle$, the probabilities of measurement outcomes labeled by the index $i$ are 
\begin{equation}
p_{i}= \langle\psi|\mathcal{P}_{i}|\psi\rangle\label{bornprob}
\end{equation}
for the set $\{\mathcal{P}_{i}\}$ of operators satisfying the condition of completeness,
\begin{equation}
\sum_{i}\mathcal{P}_{i}=\mathcal{I}\label{completeness}
\end{equation}
for identity operator $\mathcal{I}$, and the conditions of idempotency and orthogonality,
\begin{equation}
\mathcal{P}_{i}\mathcal{P}_{j}=\delta_{ij}\mathcal{P}_{j}.\label{idemortho}
\end{equation}
To demonstrate the Born rule's failure in the setting described above, Page first identifies a set of probabilities that we might have interest in computing and then proves that no complete set of orthogonal projection operators exists whose expectation values yield these probabilities. 

This demonstration relies on the imposition of two additional assumptions, his no extra vision principle and his probability symmetry principle. The former principle states that the probabilities of measurement outcomes represented nowhere in the universal quantum state are zero for all observers. In other words, a local observer should not have access to any information that a hypothetical superobserver cannot procure. I view the no extra vision principle as a fact of quantum theory's formalism. Since I only consider standard quantum theory in this paper, the reader may always adopt this attitude towards the no extra vision principle; nevertheless, I continue to refer to this fact as the no extra vision principle for consistency with Page's treatment in reference \cite{DNP}. There he contemplates theories that need not necessarily respect this principle. The probability symmetry principle states that the probabilities of measurement outcomes represented in equal numbers in the universal quantum state are equal for all observers. In other words, a local observer should not have access to any information different from that her several copies can procure. For the moment I merely wish to convey these two principles' conceptual content. When I invoke them below, I make their meanings and implications mathematically precise.

With the plan of Page's proof set forth, consider the case in which there exist two copies of an exactly identical system at different spacetime locations in our universe. I denote these systems as $\Sigma$ and $\Omega$. Of course, this distinction has no physical import: a local observer in either system knows neither which system she inhabits nor how many such systems she could inhabit. Suppose that the Hilbert spaces $\mathcal{H}_{\Sigma}$ and $\mathcal{H}_{\Omega}$ associated to these two respective systems are $2$-dimensional. Let $\{|1\rangle_{\Sigma},|2\rangle_{\Sigma}\}$ and $\{|1\rangle_{\Omega},|2\rangle_{\Omega}\}$ be orthonormal bases of states spanning these respective Hilbert spaces. For notational simplicity I denote all states of the form $|\xi\rangle_{\Sigma}\otimes|\eta\rangle_{\Omega}$ as $|\xi\eta\rangle$. I define two sets of projection operators for these two systems:
\begin{subequations}
\begin{eqnarray}
\{\mathcal{P}_{1}^{\Sigma}=|1\rangle_{\Sigma\Sigma}\langle1|,\,\mathcal{P}_{2}^{\Sigma}=|2\rangle_{\Sigma\Sigma}\langle2|\}\\
\{\mathcal{P}_{1}^{\Omega}=|1\rangle_{\Omega\Omega}\langle1|,\,\mathcal{P}_{2}^{\Omega}=|2\rangle_{\Omega\Omega}\langle2|\}
\end{eqnarray}
\end{subequations}
An observer in either system $\Sigma$ or system $\Omega$ might wish to compute the probabilities $p_{1}$ and $p_{2}$ of respectively measuring the outcomes corresponding to the states $|1\rangle$ and $|2\rangle$. To employ the Born rule she must find a complete set $\{\mathcal{P}_{1},\mathcal{P}_{2}\}$ of orthogonal projection operators  whose respective expectation values yield the probabilities $p_{1}$ and $p_{2}$. I now show by elucidating a contradiction that no such pair of operators exists.

As this demonstration relies on the imposition of the no extra vision and probability symmetry principles, I now make precise their meanings in this setting. The $4$-dimensional Hilbert space $\mathcal{H}=\mathcal{H}_{\Sigma}\otimes\mathcal{H}_{\Omega}$ of the two copies contains only two states (up to global phase) for which the potential measurement outcome corresponding to either the state $|1\rangle$ or the state $|2\rangle$ is nowhere represented: in the state $|11\rangle$ the measurement outcome corresponding to the state $|2\rangle$ has zero amplitude, and in the state $|22\rangle$ the measurement outcome corresponding to the state $|1\rangle$ has zero amplitude. The no extra vision principle then dictates that the probability $p_{1}$ of measuring the outcome corresponding to the state $|1\rangle$ in the state $|22\rangle$ must be zero and that the probability $p_{2}$ of measuring the outcome corresponding to the state $|2\rangle$ in the state $|11\rangle$ must be zero. 

The Hilbert space $\mathcal{H}$ of the two copies also contains a two real parameter family of states (up to global phase) for which the potential measurement outcomes corresponding to the states $|1\rangle$ and $|2\rangle$ are represented in equal numbers: the states $\beta_{12}|12\rangle+\beta_{21}|21\rangle$ subject to the normalization $|\beta_{12}|^{2}+|\beta_{21}|^{2}=1$. The probability symmetry principle then dictates that the probability $p_{1}$ of measuring the outcome corresponding to the state $|1\rangle$ must equal the probability $p_{2}$ of measuring the outcome corresponding to the state $|2\rangle$ in all states belonging to this family. Since the probabilities $p_{1}$ and $p_{2}$ must add to unity, both $p_{1}$ and $p_{2}$ must equal $\frac{1}{2}$.

I now proceed with the proof that no appropriate pair $\{\mathcal{P}_{1},\mathcal{P}_{2}\}$ of operators exists. First, applying the no extra vision principle to the operators $\mathcal{P}_{1}$ and $\mathcal{P}_{2}$, I require that 
\begin{subequations}
\begin{eqnarray}
\langle22|\mathcal{P}_{1}|22\rangle=0\label{nevp22a}\\ \langle11|\mathcal{P}_{2}|11\rangle=0\label{nevp22b}
\end{eqnarray}
\end{subequations}
My definition for the set $\{\mathcal{P}_{1},\mathcal{P}_{2}\}$ of operators---in particular, that their respective expectation values give the probabilities $p_{1}$ and $p_{2}$---requires  that
\begin{subequations}
\begin{eqnarray}
\langle11|\mathcal{P}_{1}|11\rangle=1\\ \langle22|\mathcal{P}_{2}|22\rangle=1\label{probdef22}
\end{eqnarray}
\end{subequations}
The conditions \eqref{nevp22a} and \eqref{nevp22b} required by the no extra vision principle also follow from this aspect of my definition for the set $\{\mathcal{P}_{1},\mathcal{P}_{2}\}$ of operators. Together these conditions dictate that 
\begin{subequations}
\begin{eqnarray}
\mathcal{P}_{1}=|11\rangle\langle11|+\mathcal{P}'_{1}\quad\mathrm{with}\quad\langle11|\mathcal{P}'_{1}|11\rangle=0\quad\mathrm{and}\quad\langle22|\mathcal{P}'_{1}|22\rangle=0\\
\mathcal{P}_{2}=|22\rangle\langle22|+\mathcal{P}'_{2}\quad\mathrm{with}\quad\langle11|\mathcal{P}'_{2}|11\rangle=0\quad\mathrm{and}\quad\langle22|\mathcal{P}'_{2}|22\rangle=0
\end{eqnarray}
\end{subequations}
Next, I insist that the pair $\{\mathcal{P}_{1},\mathcal{P}_{2}\}$ of operators satisfy the conditions expressed in \eqref{completeness} and \eqref{idemortho}. Idempotency and orthogonality imply that
\begin{equation}
\mathcal{P}'_{i}\mathcal{P}'_{j}=\delta_{ij}\mathcal{P}'_{j}.
\end{equation}
With completeness I determine that the pair $\{\mathcal{P}'_{1},\mathcal{P}'_{2}\}$ of operators must itself form a complete set of orthogonal projection operators for the Hilbert subspace spanned by the orthonormal basis $\{|12\rangle,|21\rangle\}$. Accordingly, the most general form that these operators can assume is as follows:
\begin{subequations}
\begin{eqnarray}
\mathcal{P}'_{1}=|\bar{\psi}_{12}\rangle\langle\bar{\psi}_{12}|\quad\mathrm{with}\quad|\bar{\psi}_{12}\rangle=+\cos{\theta}|12\rangle+\sin{\theta}e^{+i\phi}|21\rangle\label{state22bar}\\
\mathcal{P}'_{2}=|\tilde{\psi}_{12}\rangle\langle\tilde{\psi}_{12}|\quad\mathrm{with}\quad|\tilde{\psi}_{12}\rangle=-\sin{\theta}e^{-i\phi}|12\rangle+\cos{\theta}|21\rangle\label{state22tilde}
\end{eqnarray}
\end{subequations}
for arbitrary real parameters $\theta$ and $\phi$.

Now, I note that for the state $|\bar{\psi}_{12}\rangle$
\begin{equation}
p_{1}=\langle\bar{\psi}_{12}|\mathcal{P}_{1}|\bar{\psi}_{12}\rangle=1\quad\mathrm{and}\quad p_{2}=\langle\bar{\psi}_{12}|\mathcal{P}_{2}|\bar{\psi}_{12}\rangle=0\label{psp22violationa}
\end{equation}
and that for the state $|\tilde{\psi}_{12}\rangle$
\begin{equation}
p_{1}=\langle\tilde{\psi}_{12}|\mathcal{P}_{1}|\tilde{\psi}_{12}\rangle=0\quad\mathrm{and}\quad p_{2}=\langle\tilde{\psi}_{12}|\mathcal{P}_{2}|\tilde{\psi}_{12}\rangle=1.\label{psp22violationb}
\end{equation}
The contents of \eqref{psp22violationa} and \eqref{psp22violationb} represent violations of the probability symmetry principle, which in this case requires that
\begin{equation}
p_{1}=\langle\psi_{12}|\mathcal{P}_{1}|\psi_{12}\rangle=\frac{1}{2}\quad\mathrm{and}\quad p_{2}=\langle\psi_{12}|\mathcal{P}_{2}|\psi_{12}\rangle=\frac{1}{2}\label{psp22}
\end{equation}
for any normalized state $|\psi_{12}\rangle$ of the form $\beta_{12}|12\rangle+\beta_{21}|21\rangle$. Thus, there exists no complete set of orthogonal projection operators for which the Born rule produces all relevant probabilities as appropriate expectation values. Page admits that one might question the validity of the probability symmetry principle since the above contradiction stems from its enforcement. I, however, agree with him that this principle is not only plausible, but also desirable.

\section{Generalized Measurements to the Rescue}

I now demonstrate that the dilemma just described derives from the adoption of too restrictive a definition for the Born rule. By allowing for generalized measurements---in particular, employing positive operator valued measures, not just projection valued measures---the dilemma disappears. In other words, accepting Page's theoretical setting without modification, modern quantum measurement theory completely accommodates the types of observations that Page wishes to consider. Indeed, as we shall discover, Page, presumably unknowingly, puts us on the path to this Born rule rescue. 

Before making this demonstration explicit, I briefly justify the use of generalization measurements in the Born rule in $\S 2.1$. Although developed in the first three post war decades---see, for instance, reference \cite{KK}---the formalism of generalized measurements remains largely unknown outside of the quantum information and computation communities. In a paper on quantum cosmological issues, I thus find warranted an introductory discussion. Next in $\S 2.2$ I construct generalized measurement operators that through the Born rule yield all of the probabilities that Page desires to compute. I exhibit these operators in a series of three steps, gradually generalizing to Page's most encompassing scenario.

I must first note that Page acknowledges the possibility of employing operators more general than projection valued measures in the Born rule. In fact, for his first five alternative theories, he explicitly constructs sets $\{\mathcal{O}_{i}^{(\psi)}\}$ of ``observation" operators for computing probabilities as expectation values. These operators satisfy the positivity condition
\begin{equation}
\langle\psi|\mathcal{O}_{i}^{(\psi)}|\psi\rangle\geq 0
\end{equation}
and the completeness condition
\begin{equation}
\sum_{i}\langle\psi|\mathcal{O}_{i}^{(\psi)}|\psi\rangle=1
\end{equation}
for any state $|\psi\rangle$; however, he generically allows for these operators to depend on the state $|\psi\rangle$, which motivates my giving them such an explicit label \cite{DNP}. Operators with such dependence do not strictly count as generalized measurements within the context of quantum measurement theory; accordingly, I too consider them modifications of the Born rule. As I shall show, though, not all of these ``observation" operators fall into this category.

\subsection{Justifying Generalized Measurements}

Contemporary practitioners of quantum information science would state quantum theory's measurement postulate as follows. (See, for instance, the standard reference \cite{MAN&ILC}.) Let $\{\mathcal{M}_{ip}\}$ be a set of operators acting on the Hilbert space $\mathcal{H}$ associated to a quantum mechanical system. If these operators satisfy the completeness condition
\begin{equation}
\sum_{i}\sum_{p}\mathcal{M}_{ip}^{\dag}\mathcal{M}_{ip}=\mathcal{I},\label{gencompleteness}
\end{equation}
with $\mathcal{I}$ the identity operator on the Hilbert space $\mathcal{H}$, then this set of operators constitutes a generalized measurement with potential outcomes labeled by the index $i$. The optional index $p$ allows for the transformation of a pure state into a mixed state upon measurement as will become evident in \eqref{poststatepure} below. With the premeasurement state of the system described by the density operator $\rho$, the probability $p_{i}$ of measuring the outcome $i$ is 
\begin{equation}
\sum_{p}\mathrm{Tr}(\mathcal{M}_{ip}\rho\mathcal{M}_{ip}^{\dag}),\label{genborn},
\end{equation}
which is clearly the Born rule for a generalized measurement. After a measurement resulting in outcome $i$, the state is updated to
\begin{equation}
\frac{\sum_{p}\mathcal{M}_{ip}\rho\mathcal{M}_{ip}^{\dag}}{\sum_{p}\mathrm{Tr}(\mathcal{M}_{ip}\rho\mathcal{M}_{ip}^{\dag})}.
\end{equation}
When the density operator is pure---that is, $\rho=|\psi\rangle\langle\psi|$ for some state $|\psi\rangle$ contained in the Hilbert space $\mathcal{H}$---the two previous formulas reduce respectively to
\begin{equation}
\sum_{p}\langle\psi|\mathcal{M}_{ip}^{\dag}\mathcal{M}_{ip}|\psi\rangle\label{genbornpure}
\end{equation}
and
\begin{equation}
\frac{\sum_{p}\mathcal{M}_{ip}|\psi\rangle\langle\psi|\mathcal{M}_{ip}^{\dag}}{\sum_{p}\langle\psi|\mathcal{M}_{ip}^{\dag}\mathcal{M}_{ip}|\psi\rangle}.\label{poststatepure}
\end{equation}
The completeness condition \eqref{gencompleteness} allows for the quantities $p_{i}$ to be interpreted as probabilities computed according to the Born rule \eqref{genborn} since this condition ensures that the probabilities sum to unity. Indeed, this consequence of the completeness condition \eqref{gencompleteness} largely underpins my defining the set $\{\mathcal{M}_{ip}\}$ of operators as a quantum mechanical measurement. At a more formal level Gleason's theorem dictates this formalism as the most general that is compatible with the standard probability interpretation of quantum theory \cite{PB&PJL&PM}.

The generality of the above measurement postulate also stems from the diverse array of manipulations that theorists have contemplated for and that experimentalists have implemented on quantum mechanical systems beyond those envisioned by von Neumann. Of course this formalism encompasses projective measurements: if the measurement operators $\mathcal{M}_{i}$ satisfy the conditions of idempotency and orthogonality, namely $\mathcal{M}_{i}\mathcal{M}_{j}=\delta_{ij}\mathcal{M}_{j}$, then $\{\mathcal{M}_{i}\}$ forms a complete set of orthogonal projection operators. As physicists have discovered, however, projective measurements alone do not suffice for all experimental situations.\footnote{I must qualify this statement. There exists a method for augmenting projective measurements to make them as comprehensive as generalized measurements. By allowing for the observer to incorporate an ancillary system and to perform arbitrary unitary operations on the object plus ancillary system, she can simulate any generalized measurement with only projective measurements. This result is Neumark's theorem. See, for instance, reference \cite{AP}. For the quantum cosmological scenario at hand---and presumably many other circumstances---this method is not available: a local observer does not have access to a system ancillary to the multiple copies of exactly identical systems. The situation is not problematic: we can directly implement the necessary generalized measurements. I have of course implicitly assumed that in principle we can perform any generalized measurement. While this assumption is probably not rigorously provable, I nevertheless adopt it as a working hypothesis for two reasons. First, the relevant literature supports this supposition; see, for instance, references \cite{EL}. Second, Page appears to adhere to this attitude in reference \cite{DNP} with regard to projective measurements.} We require generalized measurements to glean information from quantum mechanical systems in those circumstances where projective measurements fall short. Most conspicuously, generalized measurements permit us to obtain at once information about both of a pair of complementary quantities; though, the amount of information obtained about either of the pair of complementary quantities from a generalized measurement is less than the amount of information obtained about one of the pair of complementary quantities from a projective measurement. This feature suggests the categorization of generalized measurements as either orthogonal or nonorthogonal: the former if projective, the latter if not projective.\footnote{The terminologies sharp or unsharp and ideal or nonideal are also in use.} Furthermore, owing to their utility for manipulating and interrogating quantum mechanical systems, experimentalists now commonly implement generalized measurements. For instance, the optimal technique for distinguishing among a set of nonorthogonal quantum states---an important task in quantum computational protocols---involves the implementation of a nonorthogonal measurement \cite{AP2}. More particularly, many observations---from the phase conjugate to particle number \cite{PB&MG&PJL2} to ascertaining spacetime coordinates \cite{PB&MG&PJL,MT}---are implementations of nonorthogonal measurements. Indeed, any measurement that is not repeatable in the projective sense is formally a nonorthogonal measurement.

To any generalized measurement defined by the set $\{\mathcal{M}_{ip}\}$ of operators, we may associate a positive operator valued measure (POVM), a set $\{\mathcal{Q}_{i}\}$ of operators acting on the Hilbert space $\mathcal{H}$ that satisfies the positivity condition
\begin{equation}
\langle\psi|\mathcal{Q}_{i}|\psi\rangle\geq0\label{positivity}
\end{equation}
for each POVM element $\mathcal{Q}_{i}$ and any state $|\psi\rangle$ and that forms a resolution of the identity,
\begin{equation}
\sum_{i}\mathcal{Q}_{i}=\mathcal{I}.\label{resolution}
\end{equation}
(When the generalized measurement is orthogonal, we refer to the set $\{\mathcal{Q}_{i}\}$ of operators as a projection valued measure.) We make the identification
\begin{equation}
\mathcal{Q}_{i}=\sum_{p}\mathcal{M}_{ip}^{\dag}\mathcal{M}_{ip}\label{identification}
\end{equation}
associating each POVM element $\mathcal{Q}_{i}$ with the respective sum of products of the measurement operators $\mathcal{M}_{ip}$. Positivity of the POVM elements, namely \eqref{positivity}, follows from the structure of the products $\mathcal{M}_{ip}^{\dag}\mathcal{M}_{ip}$ while completeness of the POVM elements, namely \eqref{resolution}, is equivalent to \eqref{gencompleteness}. The probability $p_{i}$ of measuring the outcome $i$ is now
\begin{equation}
\mathrm{Tr}(\rho\mathcal{Q}_{i})\label{povmborn}
\end{equation}
when the premeasurement state of the system is described by the density operator $\rho$ and
\begin{equation}
\langle\psi|\mathcal{Q}_{i}|\psi\rangle\label{povmbornpure}
\end{equation}
when the premeasurement state of the system is described by the pure state $|\psi\rangle$. Clearly, \eqref{povmborn} and \eqref{povmbornpure} are respectively equivalent to \eqref{genborn} and \eqref{genbornpure}. The formalism of POVMs thus provides a convenient and compact method for extracting the statistics of a generalized measurement, particularly when the postmeasurement state is not of interest.

When passing to the POVM associated with a generalized measurement, we forfeit information regarding the postmeasurement state  for the following reason. The identification \eqref{identification} does not define a one-to-one map from generalized measurements operators $\mathcal{M}_{ip}$ to POVM elements $\mathcal{Q}_{i}$; rather, the identification \eqref{identification} generally defines a many-to-one map from generalized measurements operators $\mathcal{M}_{ip}$ to POVM elements $\mathcal{Q}_{i}$. We may always write a generalized measurement operator $\mathcal{M}_{ip}$ in its polar decomposition:
\begin{equation}
\mathcal{M}_{ip}=\mathcal{U}_{ip}\sqrt{\mathcal{Q}_{i}}
\end{equation}
for a particular unitary operator $\mathcal{U}_{ip}$. Then, in forming the sum $\sum_{p}\mathcal{M}_{ip}^{\dag}\mathcal{M}_{ip}$ of products  of measurement operators to define the associated POVM element $\mathcal{Q}_{i}$, the unitary operators combine to yield the identity operator. We thus lose the information contained in the set $\{\mathcal{U}_{ip}\}$ of unitary operators that distinguishes two distinct sets $\{\mathcal{M}_{ip}\}$ and $\{\mathcal{M}'_{jq}\}$ of measurement operators having the same associated POVM. When constructing generalized measurements below, I exhibit only the POVMs as we are primarily concerned with the statistics of outcomes. This tack accords with Page's approach of attempting to find appropriate projection valued measures.

\subsection{Constructing Generalized Measurements}

\subsubsection{The Simplest Nontrivial Case}
Suppose, as in $\S1$, that the universe contains two copies of an exactly identical system at different spacetime locations each of which has associated to it a $2$-dimensional Hilbert space.\footnote{As the reader will notice momentarily, I here change the notation from $\S1$ to make the ensuing progression of generalizations more manifest.} The most general pure state of these two systems is 
\begin{equation}
|\psi^{(2,2)}\rangle=\sum_{m_{\Sigma_{1}}=1}^{2}\sum_{m_{\Sigma_{2}}=1}^{2}\beta_{m_{\Sigma_{1}}m_{\Sigma_{2}}}|m_{\Sigma_{1}}m_{\Sigma_{2}}\rangle\label{state22}
\end{equation}
subject to the normalization condition
\begin{equation}
\sum_{m_{\Sigma_{1}}=1}^{2}\sum_{m_{\Sigma_{2}}=1}^{2}|\beta_{m_{\Sigma_{1}}m_{\Sigma_{2}}}|^{2}=1.
\end{equation}

I construct a POVM satisfying the no extra vision and probability symmetry principles the expectation values of whose elements in the state $|\psi^{(2,2)}\rangle$ yield the probabilities $p_{1}^{(2,2)}$ and $p_{2}^{(2,2)}$ of respectively measuring the outcomes corresponding to the states $|1\rangle$ and $|2\rangle$. Consider the two operators
\begin{subequations}
\begin{eqnarray}
\mathcal{Q}_{1}^{(2,2)}=\mathcal{P}_{1}^{\Sigma_{1}}\otimes\mathcal{P}_{1}^{\Sigma_{2}}+\frac{1}{2}\left(\mathcal{P}_{1}^{\Sigma_{1}}\otimes\mathcal{P}_{2}^{\Sigma_{2}}+\mathcal{P}_{2}^{\Sigma_{1}}\otimes\mathcal{P}_{1}^{\Sigma_{2}}\right)\label{povm22a}\\ \mathcal{Q}_{2}^{(2,2)}=\mathcal{P}_{2}^{\Sigma_{1}}\otimes\mathcal{P}_{2}^{\Sigma_{2}}+\frac{1}{2}\left(\mathcal{P}_{1}^{\Sigma_{1}}\otimes\mathcal{P}_{2}^{\Sigma_{2}}+\mathcal{P}_{2}^{\Sigma_{1}}\otimes\mathcal{P}_{1}^{\Sigma_{2}}\right)\label{povm22b}
\end{eqnarray}
\end{subequations}
These two operators constitute a POVM: each POVM element meets the positivity condition \eqref{positivity},
\begin{subequations}
\begin{eqnarray}
\langle\psi^{(2,2)}|\mathcal{Q}_{1}^{(2,2)}|\psi^{(2,2)}\rangle=|\beta_{11}|^{2}+\frac{1}{2}\left(|\beta_{12}|^{2}|+\beta_{21}|^{2}\right)\geq 0\label{ev22a}\\
\langle\psi^{(2,2)}|\mathcal{Q}_{2}^{(2,2)}|\psi^{(2,2)}\rangle=|\beta_{22}|^{2}+\frac{1}{2}\left(|\beta_{12}|^{2}|+\beta_{21}|^{2}\right)\geq 0\label{ev22b}
\end{eqnarray}
\end{subequations}
and together the POVM elements meet the completeness condition \eqref{resolution},
\begin{equation}
\mathcal{Q}_{1}^{(2,2)}+\mathcal{Q}_{2}^{(2,2)}=\mathcal{P}_{1}^{\Sigma_{1}}\otimes\mathcal{P}_{1}^{\Sigma_{2}}+\mathcal{P}_{1}^{\Sigma_{1}}\otimes\mathcal{P}_{2}^{\Sigma_{2}}+\mathcal{P}_{2}^{\Sigma_{1}}\otimes\mathcal{P}_{1}^{\Sigma_{2}}+\mathcal{P}_{2}^{\Sigma_{1}}\otimes\mathcal{P}_{2}^{\Sigma_{2}}=\mathcal{I}^{(2,2)}.
\end{equation}
Of course, as the reader may readily confirm, they fail to meet the idempotency and orthogonality conditions.

Now, I observe that 
\begin{subequations}
\begin{eqnarray}
\langle22|\mathcal{Q}_{1}^{(2,2)}|22\rangle=0\\ \langle11|\mathcal{Q}_{2}^{(2,2)}|11\rangle=0
\end{eqnarray}
\end{subequations}
Thus, these two operators respect the no extra vision principle: they satisfy the conditions expressed in \eqref{nevp22a} and \eqref{nevp22b}. Furthermore, I observe that, for the state $|\bar{\psi}_{12}^{(2,2)}\rangle$ given in \eqref{state22bar},
\begin{subequations}
\begin{eqnarray}
p_{1}^{(2,2)}=\langle\bar{\psi}_{12}^{(2,2)}|\mathcal{Q}_{1}^{(2,2)}|\bar{\psi}_{12}^{(2,2)}\rangle=\frac{1}{2}\\
p_{2}^{(2,2)}=\langle\bar{\psi}_{12}^{(2,2)}|\mathcal{Q}_{2}^{(2,2)}|\bar{\psi}_{12}^{(2,2)}\rangle=\frac{1}{2}
\end{eqnarray}
\end{subequations}
and that, for the state $|\tilde{\psi}_{12}^{(2,2)}\rangle$ given in \eqref{state22tilde},
\begin{subequations}
\begin{eqnarray}
p_{1}^{(2,2)}=\langle\tilde{\psi}_{12}^{(2,2)}|\mathcal{Q}_{1}^{(2,2)}|\tilde{\psi}_{12}^{(2,2)}\rangle=\frac{1}{2}\\
p_{2}^{(2,2)}=\langle\tilde{\psi}_{12}^{(2,2)}|\mathcal{Q}_{2}^{(2,2)}|\tilde{\psi}_{12}^{(2,2)}\rangle=\frac{1}{2}
\end{eqnarray}
\end{subequations}
in complete accord with the probability symmetry principle. Indeed, for any normalized state $|\psi_{12}^{(2,2)}\rangle$ of the form $\beta_{12}|12\rangle+\beta_{21}|21\rangle$,
\begin{subequations}
\begin{eqnarray}
p_{1}^{(2,2)}=\langle\psi_{12}^{(2,2)}|\mathcal{Q}_{1}^{(2,2)}|\psi_{12}^{(2,2)}\rangle=\frac{1}{2}\\
p_{2}^{(2,2)}=\langle\psi_{12}^{(2,2)}|\mathcal{Q}_{2}^{(2,2)}|\psi_{12}^{(2,2)}\rangle=\frac{1}{2}
\end{eqnarray}
\end{subequations}
Finally, consider the expectation values of $\mathcal{Q}_{1}^{(2,2)}$ and of $\mathcal{Q}_{2}^{(2,2)}$ in the state $|\psi^{(2,2)}\rangle$ given respectively in \eqref{ev22a} and \eqref{ev22b}. Clearly, these are the values that we wish to assign as the probabilities $p_{1}^{(2,2)}$ and $p_{2}^{(2,2)}$ of respectively measuring the outcomes corresponding to the states $|1\rangle$ and $|2\rangle$.

Before considering the first generalization of the above POVM, I express the operators $\mathcal{Q}_{1}^{(2,2)}$ and $\mathcal{Q}_{2}^{(2,2)}$ in a slightly different form. Note that
\begin{subequations}
\begin{eqnarray}
\mathcal{Q}_{1}^{(2,2)}=\frac{\sum_{L=1}^{2}\mathscr{P}_{1}^{\Sigma_{L}}}{\langle\sum_{i}\sum_{L=1}^{2}\mathscr{P}_{i}^{\Sigma_{L}}\rangle}\label{oo22a}\\
\mathcal{Q}_{2}^{(2,2)}=\frac{\sum_{L=1}^{2}\mathscr{P}_{2}^{\Sigma_{L}}}{\langle\sum_{i}\sum_{L=1}^{2}\mathscr{P}_{i}^{\Sigma_{L}}\rangle}\label{oo22b}
\end{eqnarray}
\end{subequations}
where
\begin{subequations}
\begin{eqnarray}
\mathscr{P}_{i}^{\Sigma_{1}}=\mathcal{P}_{i}^{\Sigma_{1}}\otimes\mathcal{I}^{\Sigma_{2}}\\ \mathscr{P}_{i}^{\Sigma_{2}}=\mathcal{I}^{\Sigma_{1}}\otimes\mathcal{P}_{i}^{\Sigma_{2}}
\end{eqnarray}
\end{subequations}
and the expectation value is taken in any state of the form \eqref{state22}. For, with
\begin{subequations}
\begin{eqnarray}
\mathcal{I}^{\Sigma_{1}}=\mathcal{P}_{1}^{\Sigma_{1}}+\mathcal{P}_{2}^{\Sigma_{1}}\\
\mathcal{I}^{\Sigma_{2}}=\mathcal{P}_{1}^{\Sigma_{2}}+\mathcal{P}_{2}^{\Sigma_{2}}
\end{eqnarray}
\end{subequations}
I find that
\begin{subequations}
\begin{eqnarray}
\sum_{L=1}^{2}\mathscr{P}_{1}^{\Sigma_{L}}=2\mathcal{P}_{1}^{\Sigma_{1}}\otimes\mathcal{P}_{1}^{\Sigma_{2}}+\mathcal{P}_{1}^{\Sigma_{1}}\otimes\mathcal{P}_{2}^{\Sigma_{2}}+\mathcal{P}_{2}^{\Sigma_{1}}\otimes\mathcal{P}_{1}^{\Sigma_{2}}\\
\sum_{L=1}^{2}\mathscr{P}_{2}^{\Sigma_{L}}=2\mathcal{P}_{2}^{\Sigma_{1}}\otimes\mathcal{P}_{2}^{\Sigma_{2}}+\mathcal{P}_{1}^{\Sigma_{1}}\otimes\mathcal{P}_{2}^{\Sigma_{2}}+\mathcal{P}_{2}^{\Sigma_{1}}\otimes\mathcal{P}_{1}^{\Sigma_{2}}
\end{eqnarray}
\end{subequations}
and that
\begin{equation}
\sum_{i}\sum_{L=1}^{2}\mathscr{P}_{i}^{\Sigma_{L}}=2\mathcal{I}^{(2,2)}.
\end{equation}
The expressions \eqref{oo22a} and \eqref{oo22b} for $\mathcal{Q}_{1}^{(2,2)}$ and $\mathcal{Q}_{2}^{(2,2)}$ are precisely those for the ``observation" operators of Page's third theory for the case of two copies of an exactly identical system each of which has associated to it a $2$-dimensional Hilbert space. As we shall see momentarily, Page thus provides the first generalization of the above POVM.

\subsubsection{An Important Intermediate Case}
Suppose now that the universe contains $N$ copies of an exactly identical system at different spacetime locations each of which has associated to it an $l$-dimensional Hilbert space. The most general pure state for these $N$ systems is
\begin{equation}
|\psi^{(N,l)}\rangle=\sum_{m_{\Sigma_{1}}=1}^{l}\cdots\sum_{m_{\Sigma_{N}}=1}^{l}\beta_{m_{\Sigma_{1}}\cdots m_{\Sigma_{N}}}|m_{\Sigma_{1}}\cdots m_{\Sigma_{N}}\rangle\label{stateNl}
\end{equation}
subject to the normalization condition
\begin{equation}
\sum_{m_{\Sigma_{1}}=1}^{l}\cdots\sum_{m_{\Sigma_{N}}=1}^{l}|\beta_{m_{\Sigma_{1}}\cdots m_{\Sigma_{N}}}|^{2}=1.
\end{equation}
I formulate the no extra vision principle in $\S \mathrm{A}.1$ and the probability symmetry principle in $\S \mathrm{A}.2$ for this case. 

\paragraph{First Subcase}
I now exhibit a POVM satisfying the no extra vision and probability symmetry principles the expectation values of whose elements in the state $|\psi^{(N,l)}\rangle$ yield the probabilities $p_{i}^{(N,l)}$ of measuring the outcomes corresponding to the states $|i\rangle$. Consider the set $\{\mathcal{Q}_{i}^{(N,l)}\}$ of operators defined by 
\begin{equation}
\mathcal{Q}_{i}^{(N,l)}=\frac{\sum_{L=1}^{N}\mathscr{P}_{i}^{\Sigma_{L}}}{\langle\sum_{j}\sum_{L=1}^{N}\mathscr{P}_{j}^{\Sigma_{L}}\rangle}\label{povmNl}
\end{equation}
where
\begin{equation}
\mathscr{P}_{i}^{\Sigma_{L}}=\mathcal{I}^{\Sigma_{1}}\otimes\cdots\otimes\mathcal{I}^{\Sigma_{L-1}}\otimes\mathcal{P}_{i}^{\Sigma_{L}}\otimes\mathcal{I}^{\Sigma_{L+1}}\otimes\cdots\otimes\mathcal{I}^{\Sigma_{N}},\label{projNl}
\end{equation}
and the expectation value is taken in any state of the form \eqref{stateNl}. The operators $\mathcal{Q}_{i}^{(N,l)}$ are the ``observation" operators of Page's third theory \cite{DNP}. Noting that
\begin{equation}
\sum_{i}\sum_{L=1}^{N}\mathscr{P}_{i}^{\Sigma_{L}}=N\mathcal{I}^{(N,l)},\label{resolutionNl}
\end{equation}
I may write
\begin{equation}
\mathcal{Q}_{i}^{(N,l)}=\frac{1}{N}\sum_{L=1}^{N}\mathscr{P}_{i}^{\Sigma_{L}}.
\end{equation}
In $\S \mathrm{A}.3.1$ I prove that this set of operators constitutes a POVM possessing the desired properties.

My analysis demonstrates that Page's third theory does not represent a modification of the Born rule but rather a deployment of a POVM. Hartle and Srednicki have claimed that this third theory represents the use of a xerographic distribution in conjunction with (Page's definition of) the Born rule \cite{JH&MS}. (These authors made this statement in an earlier unpublished version of reference \cite{JH&MS}.) No doubt we could construct the operators \eqref{povmNl} in this fashion. I contend, however, that recognizing these operators as constituting a POVM represents a simpler interpretation: no extra element need be appended to the formalism of quantum measurement theory. 

Page develops his third theory on the basis of a volume weighting measure \cite{DNP}. Ultimately, he disfavors this theory, arguing that it possibly suffers from a Boltzmann brain problem \cite{DNP,DNP2}. As this theory is simply quantum measurement theory, I doubt that such a problem could be present, though other schemes based on a volume weighting measure may well possess such issues. Now, I have not shown that the POVM \eqref{povmNl} is the unique set of operators within the confines of quantum measurement theory; for, there could exist more than one POVM satisfying all of the desired criteria. If a different POVM does exist, however, then it would not correspond to any of Page's other theories as the ``observation operators" for these theories are not generalized measurements of any sort. Supposing that there exists at least one different POVM, we would then be confronted with the conundrum of choosing between the two or more candidates. Presumably, we would then look for some physical principle---potentially akin to Page's recourse to volume weighting---to discriminate amongst the various POVMs. I am currently investigating these and related questions.

\paragraph{Second Subcase}
In the first subcase I implicitly assumed in constructing the POVM \eqref{povmNl} that, in all $N$ copies of the exactly identical system, the local observer made a measurement. This need not be the case: only $M$ of the $N$ observers might actually carry out the measurement. As any particular local observer has only local knowledge, \emph{a priori} she cannot know how many of the other observers perform the measurement.

Thus, for the case in which an unknown number $M$ of the $N$ observers carry out the observation, I exhibit a POVM satisfying the no extra vision and probability symmetry principles the expectation values of whose elements in the state $|\psi^{(N,l)}\rangle$ yield the probabilities $p_{i}^{(N,M,l)}$ of measuring the outcomes corresponding to the states $|i\rangle$. Consider the set $\{\mathcal{Q}_{i}^{(N,M,l)}\}$ of operators defined by
\begin{equation}
\mathcal{Q}_{i}^{(N,M,l)}=\frac{\sum_{L=1}^{M}\tilde{\mathscr{P}}_{i}^{\Sigma_{L}}}{\langle\sum_{j}\sum_{L=1}^{M}\tilde{\mathscr{P}}_{j}^{\Sigma_{L}}\rangle}\label{povmNN0l}
\end{equation}
where
\begin{equation}
\tilde{\mathscr{P}}_{i}^{\Sigma_{L}}=\mathcal{I}^{\Sigma_{1}}\otimes\cdots\otimes\mathcal{I}^{\Sigma_{L-1}}\otimes\mathcal{P}_{i}^{\Sigma_{L}}\otimes\mathcal{I}^{\Sigma_{L+1}}\otimes\cdots\otimes\mathcal{I}^{\Sigma_{M}}\otimes\cdots\otimes\mathcal{I}^{\Sigma_{N}},\label{projNN0l}
\end{equation}
and the expectation value is taken in any state of the form \eqref{stateNl}. I have chosen to order the $N$ identical systems so that the $M$ copies in which the local observer performs the measurement are listed first. Accordingly, in \eqref{projNN0l} for every $L$, an identity operator occupies each of last $N-M$ entries in the tensor product. Noting that
\begin{equation}
\sum_{i}\sum_{L=1}^{M}\tilde{\mathscr{P}}_{i}^{\Sigma_{L}}=M\mathcal{I}^{(N,l)},\label{resolutionNN0l}
\end{equation}
I may write
\begin{equation}
\mathcal{Q}_{i}^{(N,M,l)}=\frac{1}{M}\sum_{L=1}^{M}\tilde{\mathscr{P}}_{i}^{\Sigma_{L}}.
\end{equation}
In $\S \mathrm{A}.3.2$ I prove that this set of operators constitutes a POVM possessing the desired properties.

\subsubsection{The Most General Case}
Suppose now that the pure state  
\begin{equation}
|\psi^{(\{N\},l)}\rangle=\sum_{N=1}^{\mathscr{N}}\alpha_{N}\sum_{m_{\Sigma_{1}}=1}^{l}\cdots\sum_{m_{\Sigma_{N}}=1}^{l}\beta_{m_{\Sigma_{1}}\cdots m_{\Sigma_{N}}}|m_{\Sigma_{1}}\cdots m_{\Sigma_{N}}\rangle\label{statealephl}
\end{equation}
subject to the normalization condition
\begin{equation}
\sum_{N=1}^{\mathscr{N}}|\alpha_{N}|^{2}\sum_{m_{\Sigma_{1}}=1}^{l}\cdots\sum_{m_{\Sigma_{N}}=1}^{l}|\beta_{m_{\Sigma_{1}}\cdots m_{\Sigma_{N}}}|^{2}=1
\end{equation}
describes some part of the universe. In other words, there exists a superposition over the number $N$ of exactly identical copies of a system with $l$-dimensional Hilbert space.\footnote{The reader might be uncomfortable with the definition of the state \eqref{statealephl}. I discuss issues with this state in $\S 3.1$. For the moment we can at least make sense of this state on a purely notational level.} I formulate the no extra vision principle in $\S \mathrm{B}.1$ and the probability symmetry principle in $\S \mathrm{B}.2$ for this case. 

\paragraph{First Subcase} 
I now exhibit a POVM satisfying the no extra vision and probability symmetry principles the expectation values of whose elements in the state $|\psi^{(\{N\},l)}\rangle$ yield the probabilities $p_{i}^{(\{N\},l)}$ of measuring the outcomes corresponding to the states $|i\rangle$. Consider the set $\{\mathcal{Q}_{i}^{(\{N\},l)}\}$ of operators defined by 
\begin{equation}
\mathcal{Q}_{i}^{(\{N\},l)}=\frac{\sum_{N=1}^{\mathscr{N}}\frac{1}{N}\sum_{L=1}^{N}\mathscr{P}_{i}^{\Sigma_{L}}}{\langle\sum_{j}\sum_{N=1}^{\mathscr{N}}\frac{1}{N}\sum_{L=1}^{N}\mathscr{P}_{j}^{\Sigma_{L}}\rangle}\label{povmalephl}
\end{equation}
where
\begin{equation}
\mathscr{P}_{i}^{\Sigma_{L}}=\mathcal{I}^{\Sigma_{1}}\otimes\cdots\otimes\mathcal{I}^{\Sigma_{L-1}}\otimes\mathcal{P}^{\Sigma_{L}}\otimes\mathcal{I}^{\Sigma_{L+1}}\otimes\cdots\otimes\mathcal{I}^{\Sigma_{N}},\label{projalephl}
\end{equation}
and the expectation value is taken in any state of the form \eqref{statealephl}. Noting that
\begin{equation}
\sum_{j}\sum_{N=1}^{\mathscr{N}}\frac{1}{N}\sum_{L=1}^{N}\mathscr{P}_{j}^{\Sigma_{L}}=\mathcal{I}^{(\{N\},l)},\label{resolutionalephl}
\end{equation}
I may write
\begin{equation}
\mathcal{Q}_{i}^{(\{N\},l)}=\sum_{N=1}^{\mathscr{N}}\frac{1}{N}\sum_{L=1}^{N}\mathscr{P}_{i}^{\Sigma_{L}}.
\end{equation}
In $\S \mathrm{B}.3.1$ I prove that this set of operators constitutes a POVM possessing the desired properties.

\paragraph{Second Subcase}
In the first subcase I implicitly assumed in constructing the POVM \eqref{povmalephl} that, in every copy of the exactly identical system, the local observer made a measurement. This need not be the case: only $M_{N}$ of the $N$ observers within the branch of the universal state having $N$ exactly identical systems might actually carry out the measurement. As any particular observer has only local knowledge, \emph{a priori} she cannot know how many of the other observers in her branch of the universal state perform the measurement.

Thus, for the case in which an unknown number $M_{N}$ of the $N$ observers within the branch of the universal state having $N$ exactly identical systems carry out the observation, I exhibit a POVM satisfying the no extra vision and probability symmetry principles the expectation values of whose elements in the state $|\psi^{(\{N\},l)}\rangle$ yield the probabilities $p_{i}^{(\{N\},\{M\},l)}$ of measuring the outcomes corresponding to the states $|i\rangle$. Consider the set $\{\mathcal{Q}_{i}^{(\{N\},\{M\},l)}\}$ of operators defined by 
\begin{equation}
\mathcal{Q}_{i}^{(\{N\},\{M\},l)}=\frac{\sum_{N=1}^{\mathscr{N}}\frac{1}{M_{N}}\sum_{L=1}^{M_{N}}\tilde{\mathscr{P}}_{i}^{\Sigma_{L}}}{\langle\sum_{j}\sum_{N=1}^{\mathscr{N}}\frac{1}{M_{N}}\sum_{L=1}^{M_{N}}\tilde{\mathscr{P}}_{j}^{\Sigma_{L}}\rangle}\label{povmaleph0l}
\end{equation}
where
\begin{equation}
\tilde{\mathscr{P}}_{i}^{\Sigma_{L}}=\mathcal{I}^{\Sigma_{1}}\otimes\cdots\otimes\mathcal{I}^{\Sigma_{L-1}}\otimes\mathcal{P}_{i}^{\Sigma_{L}}\otimes\mathcal{I}^{\Sigma_{L+1}}\otimes\cdots\otimes\mathcal{I}^{\Sigma_{M}}\otimes\cdots\otimes\mathcal{I}^{\Sigma_{N}},\label{projaleph0l}
\end{equation}
and the expectation value is taken in any state of the form \eqref{statealephl}. I have chosen to order the $N$ identical systems within each branch so that the $M_{N}$ copies in which the local observer performs the measurement are listed first. Accordingly, in \eqref{projaleph0l} within each branch for every $L$, an identity operator occupies each of last $N-M_{N}$ entries in the tensor product. Noting that
\begin{equation}
\sum_{j}\sum_{N=1}^{\mathscr{N}}\frac{1}{M_{N}}\sum_{L=1}^{M_{N}}\tilde{\mathscr{P}}_{j}^{\Sigma_{L}}=\mathcal{I}^{(\{N\},l)},\label{resolutionaleph0l}
\end{equation}
I may write
\begin{equation}
\mathcal{Q}_{i}^{(\{N\},\{M\},l)}=\sum_{N=1}^{\mathscr{N}}\frac{1}{M_{N}}\sum_{L=1}^{M_{N}}\tilde{\mathscr{P}}_{i}^{\Sigma_{L}}.
\end{equation}
In $\S \mathrm{B}.3.2$ I prove that this set of operators constitutes a POVM possessing the desired properties.\newline\newline Evidently, generalized measurements rescue the Born rule for quantum cosmology, at least in the scenarios that Page contemplates: allowing myself only to update Page's definition of the Born rule, I have shown that this rule suffices for computing all of the desired probabilities. My argument may strike the reader as proverbial hair splitting over what to include in the formalism of standard quantum theory. While this interpretation of my argument is undeniably true---my conclusions rest on a redefinition of the Born rule---I contend that the chosen side of the split is a matter of importance. If we settle on Page's side, then we are forced to abandon outright or to modify substantially our most successful scientific theory. If we settle on my side, then we celebrate the agility and facility of our most successful scientific theory. In light of the consequences of the former option---notably the apparent necessity of replacing quantum theory---and of the justifications for generalized measurements---especially their now common implementation in sundry experimental procedures---I submit that Page's stated concern for the Born rule's well being is simply misplaced.

\section{Quantum Theory Strikes Again}

\subsection{Hilbert Space Dimensions in a Very Large Universe}
Consider again the most general pure state $|\psi^{(\{N\},l)}\rangle$, given in \eqref{statealephl}, that Page permits as a description of some part of the universe. The first several components of this state are notated as follows:
\begin{eqnarray}
|\psi^{(\{N\},l)}\rangle&=&\alpha_{1}\left(\beta_{1}|1\rangle+\cdots+\beta_{l}|l\rangle\right)+\alpha_{2}\left(\beta_{11}|11\rangle+\cdots+\beta_{ll}|ll\rangle\right)\nonumber\\ &&+\alpha_{3}\left(\beta_{111}|111\rangle+\cdots+\beta_{lll}|lll\rangle\right)\nonumber\\ &&+\alpha_{4}\left(\beta_{1111}|1111\rangle+\cdots+\beta_{llll}|llll\rangle\right)+\cdots.\label{statealephlexplicit}
\end{eqnarray}
For this to constitute a sensible state within the confines of standard quantum theory, each term in this superposition must represent a distinct vector in a certain Hilbert space of fixed dimension. For instance, the states labelled $|1\rangle$, $|11\rangle$, $|111\rangle$, $\ldots$ must all stand for vectors in the same Hilbert space. On a naive notational interpretation the states $|1\rangle$, $|11\rangle$, $|111\rangle$, $\ldots$ do not belong to the same Hilbert space. Rather, if the state $|1\rangle$ is an element of the Hilbert space $\mathcal{H}^{(1,l)}$, then the state $|11\rangle$ is an element of the Hilbert space $\mathcal{H}^{(2,l)}=\mathcal{H}^{(1,l)}\otimes\mathcal{H}^{(1,l)}$, the state $|111\rangle$ is an element of the Hilbert space $\mathcal{H}^{(3,l)}=\mathcal{H}^{(1,l)}\otimes\mathcal{H}^{(1.l)}\otimes\mathcal{H}^{(1,l)}$, \emph{et cetera}. In this case since $\mathrm{dim}(\mathcal{H}^{(1,l)})=l$, $\mathrm{dim}(\mathcal{H}^{(2,l)})=l^{2}$, $\mathrm{dim}(\mathcal{H}^{(3,l)})=l^{3}$, \emph{et cetera}.

For this interpretation to be in error, we must revise or suppress our understanding of the notation. How can we make sense of the states $|1\rangle$, $|11\rangle$, $|111\rangle$, $\ldots$ all belonging to the same Hilbert space? Only one possibility seems plausible: the basis of states for the Hilbert space $\mathcal{H}^{(\{N\},l)}$ is composed of the elements $|1\rangle,\cdots,|l\rangle,|11\rangle,\cdots,|ll\rangle,|111\rangle,\cdots,,|lll\rangle,|1111\rangle,\cdots,|llll\rangle,\cdots$. The structure of this Hilbert space $\mathcal{H}^{(\{N\},l)}$ is, however, most curious. Formally, 
\begin{equation}
\mathcal{H}^{(\{N\},l)}=\mathcal{H}^{(1,l)}\oplus\cdots\oplus\mathcal{H}^{(L,l)}\oplus\cdots\oplus\mathcal{H}^{(N,l)}
\end{equation}
where
\begin{equation}
\mathcal{H}^{(L,l)}=\bigotimes_{K=1}^{L}\mathcal{H}_{\Sigma_{K}}^{(1,l)}.
\end{equation}
Forming the direct product of two Hilbert spaces of different dimensions is of course commonplace: any Hilbert space with one or more subspaces of differing dimensions may be represented in this manner. Forming the direct product of two Hilbert spaces of different dimensions one of which is constructed as a tensor product of copies of the other is not exactly commonplace. At a mathematical level I observe no formal obstruction to such a construction. Indeed, I implicitly interpreted the scenario of $\S 2.2.3$ in this fashion without any apparent difficulties. At a physical level I observe a conceptual obstruction to such a construction. Consider, for clarity, the special case $|1\rangle+|11\rangle$ of the state \eqref{statealephlexplicit}. The first component state in this superposition describes the state of one particular physical system, namely that of the system $\Sigma_{1}$. The second component state in this superposition describes the joint state of two of these physical systems, namely that of the two systems $\Sigma_{1}$ and $\Sigma_{2}$. Thus, the state $|1\rangle+|11\rangle$ is a superposition of two states each of which describes a different physical system. Such a state does not make physical sense in standard quantum theory where we associate different Hilbert spaces to different physical systems. I therefore submit that the state $|\psi^{(\{N\},l)}\rangle$ is not well defined. Then the state $|\psi^{(N,l)}\rangle$ would be the most general state allowed in Page's formalism. As I have previously mentioned, Page is interested in theories that differ from quantum theory. Consequently, such theories could allow for states of the form $|\psi^{(\{N\},l)}\rangle$; nevertheless, Page gives the impression of wanting to retain quantum theory's Hilbert space structure, only modifying the theory's probability rule.

\subsection{Permutation Symmetries in a Very Large Universe}
In light of the above discussion, consider now the pure state $|\psi^{(N,l)}\rangle$, given in \eqref{stateNl}, as a description of some part of the universe. The multiple copies of an exactly identical system in this envisioned universe are related by a permutation symmetry: a mere relabeling of the copies must by definition have no physical effect. We may conceptualize the indistinguishability of these multiple copies as follows: a local observer in this universe lacks a reference frame for differentiating amongst the copies.\footnote{A reference frame, classical or quantum, is any physical system that serves as a reference for the measurement of some physical quantity of another physical system. For instance, a phase-locked laser constitutes a reference frame for phase measurements, and an atomic clock constitutes a reference frame for time measurements. See reference \cite{SDB&TR&RWS} for a more extensive discussion.} The absence of such a reference frame is plain: such a reference frame would need be external to the universe. As Bartlett, Rudolph, and Spekkens have shown, the lack of a reference frame entails the existence of a superselection rule \cite{SDB&TR&RWS}.

The relevant group here is the symmetric group $S_{N}$ acting on the $N$ copies of an exactly identical system. Under the action of the symmetric group, the Hilbert space $\mathcal{H}^{(N,l)}$ spanned by the states $|\psi^{(N,l)}\rangle$ decomposes into superselection sectors, Hilbert subspaces $\mathcal{H}_{s}^{(N,l)}$ labeled by the symmetry type $s$, either antisymmetric, parasymmetric, or symmetric. Accordingly, no local observer can exhibit coherence between these superselection sectors. My recognition of the superselection rule associated with a permutation symmetry of physical systems is by no means novel: several authors noted its existence in the context of the quantum mechanics of identical particles, where, when particularly pertaining to bosons and fermions, it is referred to as the univalence superselection rule \cite{AMLM&OWG,JBH&JRT}.

This analogy to the quantum mechanics of identical particles extends further than the existence of a superselection rule for symmetry type. The states $|\psi^{(N,l)}\rangle$ are formally analogous to those of identical particles---each copy of an exactly identical system playing the role of an identical particle---except that the states $|\psi^{(N,l)}\rangle$ are considerably more general than those typically assigned to identical particles. As the reader no doubt recalls, we typically enforce an even stricter condition on the states of identical particles: the so-called symmetrization postulate. On the basis of this postulate, we limit the states of identical particles to those that are either antisymmetric or symmetric under permutations. While the superselection rule for symmetry type is a provable consequence of the lack of an appropriate reference frame \cite{AMLM&OWG, GCW&ASW&EPW, SDB&TR&RWS, MD&SDB&TR&RWS}, the symmetrization postulate does not possess this axiomatic status. Rather, we are led to its imposition by different considerations: the maintenance of locality through the cluster decomposition priniciple and the nonobservation of fundamental paraparticles.\footnote{Neither of these reasons for imposing the symmetrization postulate apply to the cosmological scenarios at hand: there is no need to demand that the multiple copies obey the cluster decomposition principle, and a local observer cannot determine the transformation properties of the state $|\psi^{(N,l)}\rangle$ under the action of $S_{N}$.}

Returning to the scenario at hand, we might wonder if any of these considerations---either those concerning the superselection rule or the symmetrization postulate---could resolve the issue that Page identifies for (his definition of) the Born rule. In other words, when accounting for the superselection rule for symmetry type on the Hilbert space $\mathcal{H}^{(N,l)}$, thereby effectively restricting the states $|\psi^{(N,l)}\rangle$ to those without coherence between superselection sectors, can we construct a complete set of orthogonal projection operators satisfying the no extra vision and probability symmetry principles that yield the probabilities $p_{i}^{(N,l)}$ as expectation values? Or, when imposing the symmetrization postulate on the Hilbert space $\mathcal{H}^{(N,l)}$, thereby restricting the states $|\psi^{(N,l)}\rangle$ to those either antisymmetric or symmetric under permutations, can we construct a complete set of orthogonal projection operators satisfying the no extra vision and probability symmetry principles that yield the probabilities $p_{i}^{(N,l)}$ as expectation values? 

I choose to explore these questions in the absence of Page's two additional principles with the hope of ascertaining two points: first, whether or not the dilemma displayed in $\S 1$ arises in the standard quantum mechanics of identical particles, and, second, whether or not the probability symmetry principle is an integral source of this dilemma. If I am able to construct a complete set $\{\mathcal{P}_{i}^{(N,l)}\}$ of orthogonal projection operators in the presence of either the superselection rule or the symmetrization postulate, then I will have pinpointed the probability symmetry principle as the culprit. If I am not able to construct a complete set $\{\mathcal{P}_{i}^{(N,l)}\}$ of orthogonal projection operators in the presence of either the superselection rule or the symmetrization principle, then I will conclude that this dilemma is characteristic of quantum theory and requires the use of generalized measurements for its alleviation.

To this end in $\S\mathrm{C}.1.1$ and $\S\mathrm{C}.2.1$ I consider again the universe containing two copies of an exactly identical system at different spacetime locations each of which has associated to it a $2$-dimensional Hilbert space. As I demonstrate there, the Born rule's purported failure occurs even for this universe when either the superselection rule or the symmetrization postulate is imposed (in the absence of the no extra vision and probability symmetry principles). Since this model universe is precisely analogous to a system of two identical spin one-half particles, for instance, I conclude that the Born rule's insufficiency (according to Page's definition) is not peculiar to the cosmological setting. In $\S 2$ I illustrated how an observer can cirumvent this apparent dilemma by employing the appropriate generalized measurements. I note further that the POVM \eqref{povm22a} and \eqref{povm22b} respects the superselection rule for symmetry type, as shown in $\S\mathrm{C}.1.2$, and that one can construct a POVM accommodating the symmetrization postulate, as shown $\S\mathrm{C}.2.2$.

\section{Concluding Comments}
Quantum theory is a remarkably general yet powerful framework that continues to amaze in its applicability to ever more esoteric settings. My work serves not only to illustrate this versatility and strength, but also to remind us to exploit these characteristics to their fullest potential. The claimed failure of quantum mechanics elucidated in $\S 1$ stemmed from ignoring the generality of quantum measurement theory. By employing this theory more completely, I readily resolved the apparent dilemma. When illuminating these novel physical situations, we must also not lose sight of quantum theory's lessons from familiar physical situations. The relevance of the cosmological setting to the argument of $\S 1$ evaporated once I drew the analogy to the quantum mechanics of identical particles; moreover, the imposition of the probability symmetry principle became unnecessary once I recalled the superselection rule for symmetry type. While I have thus succeeded in saving the Born rule, the cosmological measure problem remains outstanding. My analysis does not provide any particularly illuminating insights into this problem, except to warn against so readily jettisoning the formalism of quantum theory in developing a solution.

\section*{Acknowledgments}
First, I thank Bill Wootters for suggesting the POVM of \eqref{povm22a} and \eqref{povm22b} that instigated this research and for numerous comments on earlier versions of this paper. Second, I thank Andy Albrecht for suggesting the analogy to identical particles that sparked much of $\S 3$ and for several comments on an earlier version of this paper. Third, I thank Steve Carlip for comments on an earlier version of this paper. I also wish to thank Augusta Abrahamse, Andy Albrecht, Brandon Bozek, Damien Martin, Don Page, Dan Phillips, and Paul Teller for useful discussions. This work was supported in part by Department of Energy grant DE-FG02-91ER40674.

\appendix

\section{On ``An Important Intermediate Case"}

\subsection{Formulating the No Extra Vision Principle}
I here make precise the meaning of the no extra vision principle for the setting of $\S 2.2.2$. The $l^{N}$-dimensional Hilbert space $\mathcal{H}^{(N,l)}$ for these $N$ copies is the tensor product $\bigotimes_{L=1}^{N}\mathcal{H}_{\Sigma_{L}}^{(1,l)}$ over the $l$-dimensional Hilbert spaces $\mathcal{H}^{(1,l)}$ of the individual copies. I identify $l-1$ classes of states contained in the Hilbert space $\mathcal{H}^{(N,l)}$ for which some of the potential measurement outcomes corresponding to the states $|i\rangle$ are not represented. I notate these classes of states, labeling them by the outcomes not present, as follows:
\begin{subequations}
\begin{eqnarray}
|\psi_{\neg i}^{(N,l)}\rangle=\sum_{\substack{m_{\Sigma_{1}}=1\\ m_{\Sigma_{1}}\neq i}}^{l}\cdots\sum_{\substack{m_{\Sigma_{N}}=1\\ m_{\Sigma_{N}}\neq i}}^{l}\beta_{m_{\Sigma_{1}}\cdots m_{\Sigma_{N}}}|m_{\Sigma_{1}}\cdots m_{\Sigma_{N}}\rangle\label{nevpstateNla}\\
|\psi_{\neg i \neg j}^{(N,l)}\rangle=\sum_{\substack{m_{\Sigma_{1}}=1\\ m_{\Sigma_{1}}\neq i\\ m_{\Sigma_{1}}\neq j}}^{l}\cdots\sum_{\substack{m_{\Sigma_{N}}=1\\ m_{\Sigma_{N}}\neq i\\ m_{\Sigma_{N}}\neq j}}^{l}\beta_{m_{\Sigma_{1}}\cdots m_{\Sigma_{N}}}|m_{\Sigma_{1}}\cdots m_{\Sigma_{N}}\rangle\label{nevpstateNlb}\\
|\psi_{\neg i \neg j \neg k}^{(N,l)}\rangle=\sum_{\substack{m_{\Sigma_{1}}=1\\ m_{\Sigma_{1}}\neq i\\ m_{\Sigma_{1}}\neq j\\ m_{\Sigma_{1}}\neq k}}^{l}\cdots\sum_{\substack{m_{\Sigma_{N}}=1\\ m_{\Sigma_{N}}\neq i\\ m_{\Sigma_{N}}\neq j\\ m_{\Sigma_{N}}\neq k}}^{l}\beta_{m_{\Sigma_{1}}\cdots m_{\Sigma_{N}}}|m_{\Sigma_{1}}\cdots m_{\Sigma_{N}}\rangle\label{nevpstateNlc}\\
et\, cetera\nonumber
\end{eqnarray}
\end{subequations}
for inequivalent $i,j,k,$ \emph{et cetera} taking values in the set $\{1,\ldots,l\}$. Each successive class of states is contained within the previous class; for clarity in presenting the no extra vision principle's consequences, I find this classification most useful. This principle then dictates that the probabilities $p_{i}^{(N,l)}$ of measuring the outcomes corresponding to the states $|i\rangle$ be zero in the states \eqref{nevpstateNla}, that the respective probabilities $p_{i}^{(N,l)}$ and $p_{j}^{(N,l)}$ of respectively measuring the outcomes corresponding to the states $|i\rangle$ and $|j\rangle$ be zero in the states \eqref{nevpstateNlb}, that the probabilities $p_{i}^{(N,l)}$, $p_{j}^{(N,l)}$, and $p_{k}^{(N,l)}$ of respectively measuring the outcomes corresponding to the states $|i\rangle$, $|j\rangle$, and $|k\rangle$ be zero in the states \eqref{nevpstateNlc}, \emph{et cetera}.

\subsection{Formulating the Probability Symmetry Principle}
I now make precise the meaning of the probability symmetry principle for the setting of $\S 2.2.2$. I identify $\lfloor\frac{N}{2}\rfloor$ classes of states contained in the Hilbert space $\mathcal{H}^{(N,l)}$ for which the potential measurement outcomes corresponding to the states $|i\rangle$ and $|j\rangle$ are represented in equal numbers.\footnote{The probability symmetry principle only applies for $N\geq2$.} I notate these classes of states, labeling them by the number $n_{ij}$ of occurrences of each outcome, as follows:
\begin{subequations}
\begin{eqnarray}
|\psi_{n_{ij}=1}^{(N,l)}\rangle=\sum_{m_{\Sigma_{1}}=1}^{l}\sum_{\substack{m_{\Sigma_{2}}=1\\ m_{\Sigma_{1}}=i\Rightarrow m_{\Sigma_{2}}\neq i \\ m_{\Sigma_{1}}=j\Rightarrow m_{\Sigma_{2}}\neq j}}^{l}\cdots\beta_{m_{\Sigma_{1}}m_{\Sigma_{2}}\cdots}|m_{\Sigma_{1}}m_{\Sigma_{2}}\cdots\rangle\label{pspstateNla}\\
|\psi_{n_{ij}=2}^{(N,l)}\rangle=\sum_{m_{\Sigma_{1}}=1}^{l}\sum_{m_{\Sigma_{2}}=1}^{l}\sum_{\substack{m_{\Sigma_{3}}=1\\ m_{\Sigma_{1}}=i\wedge m_{\Sigma_{2}}=i\Rightarrow m_{\Sigma_{3}}\neq i \\ m_{\Sigma_{1}}=j\wedge m_{\Sigma_{2}}=j\Rightarrow m_{\Sigma_{3}}\neq j}}^{l}\cdots\beta_{m_{\Sigma_{1}}m_{\Sigma_{2}}m_{\Sigma_{3}}\cdots}|m_{\Sigma_{1}}m_{\Sigma_{2}}m_{\Sigma_{3}}\cdots\rangle\label{pspstateNlb}\\
|\psi_{n_{ij}=3}^{(N,l)}\rangle=\sum_{m_{\Sigma_{1}}=1}^{l}\sum_{m_{\Sigma_{2}}=1}^{l}\sum_{m_{\Sigma_{3}}=1}^{l}\sum_{\substack{m_{\Sigma_{4}}=1\\ m_{\Sigma_{1}}=i\wedge m_{\Sigma_{2}}=i\wedge m_{\Sigma_{3}}=i\Rightarrow m_{\Sigma_{4}}\neq i\\ m_{\Sigma_{1}}=j\wedge m_{\Sigma_{2}}=j\wedge m_{\Sigma_{3}}=j\Rightarrow m_{\Sigma_{4}}\neg j}}^{l}\cdots\beta_{m_{\Sigma_{1}}m_{\Sigma_{2}}m_{\Sigma_{3}}m_{\Sigma_{4}}\cdots}|m_{\Sigma_{1}}m_{\Sigma_{2}}m_{\Sigma_{3}}m_{\Sigma_{4}}\cdots\rangle\label{pspstateNlc}\\
et\,cetera\nonumber
\end{eqnarray}
\end{subequations}
for inequivalent $i,j$ taking values in the set $\{1,\ldots,l\}$.\footnote{I do not allow $n_{ij}=0$ since, for this class of states, the probability symmetry principle is equivalent to the no extra vision principle.} Each successive class of states contains the previous class; for clarity in presenting the probability symmetry principle's consequences, I find this classification most useful. This principle now dictates that, for each state in the classes above, the probabilities $p_{i}^{(N,l)}$ and $p_{j}^{(N,l)}$ of respectively measuring the outcomes corresponding to the states $|i\rangle$ and $|j\rangle$ be equal.

\subsection{Properties of the Constructed Positive Operator Valued Measures}

\subsubsection{First Subcase}
The set \eqref{povmNl} of operators constitutes a POVM: each element $\mathcal{Q}_{i}^{(N,l)}$ meets the positivity condition \eqref{positivity} since
\begin{eqnarray}
\lefteqn{\langle\psi^{(N,l)}|\mathscr{P}_{i}^{\Sigma_{L}}|\psi^{(N,l)}\rangle=}\nonumber\\ &&\sum_{m_{\Sigma_{1}}=1}^{l}\cdots\sum_{m_{\Sigma_{L-1}}=1}^{l}\sum_{m_{\Sigma_{L+1}}=1}^{l}\cdots\sum_{m_{\Sigma_{N}}=1}^{l}|\beta_{m_{\Sigma_{1}}\cdots m_{\Sigma_{L-1}}im_{\Sigma_{L+1}}\cdots m_{\Sigma_{N}}}|^{2}\geq0
\end{eqnarray}
for all $L$, and together the elements meet the completeness condition \eqref{resolution} owing to \eqref{resolutionNl}.

Now, with the probabilities $p_{i}^{(N,l)}$ given by the expectation values of $\mathcal{Q}_{i}^{(N,l)}$, I observe that this POVM respects the no extra vision principle: the expectation values of its elements $\mathcal{Q}_{i}^{(N,l)}$ are zero in the states \eqref{nevpstateNla}, the expectation values of its elements $\mathcal{Q}_{i}^{(N,l)}$ and $\mathcal{Q}_{j}^{(N,l)}$ are zero in the states \eqref{nevpstateNlb}, the expectation values of its elements $\mathcal{Q}_{i}^{(N,l)}$, $\mathcal{Q}_{j}^{(N,l)}$, and $\mathcal{Q}_{k}^{(N,l)}$ are zero in the states \eqref{nevpstateNlc}, \emph{et cetera}. As a convex combination of the projection operators $\mathscr{P}_{i}^{\Sigma_{L}}$ over all $N$ identical copies, the POVM element $\mathcal{Q}_{i}^{(N,l)}$ possesses for each copy $L$ a term containing the projection operator $\mathcal{P}_{i}^{\Sigma_{L}}$. Thus, when acting on a state with the measurement outcome corresponding to the state $|i\rangle$ nowhere represented, the POVM element $\mathcal{Q}_{i}^{(N,l)}$ annihilates the state.

Next, I observe that the above POVM respects the probability symmetry principle:  for each of the states \eqref{pspstateNla}, \eqref{pspstateNlb}, \eqref{pspstateNlc}, \emph{et cetera}, the expectation values of its elements $\mathcal{Q}_{i}^{(N,l)}$ and $\mathcal{Q}_{j}^{(N,l)}$ are equal. When acting on any of these states, both $\mathcal{Q}_{i}^{(N,l)}$ and $\mathcal{Q}_{j}^{(N,l)}$ select $n_{ij}$ times each and every component state $|m_{\Sigma_{1}}\cdots m_{\Sigma_{N}}\rangle$, yielding an expectation value of $\frac{n_{ij}}{N}$ for both POVM elements. 

Finally, consider the expectation value of $\mathcal{Q}_{i}^{(N,l)}$ in the state $|\psi^{(N,l)}\rangle$:
\begin{eqnarray}
\lefteqn{\langle\psi^{(N,l)}|\mathcal{Q}_{i}^{(N,l)}|\psi^{(N,l)}\rangle=}\nonumber\\ &&\frac{1}{N}\sum_{L=1}^{N}\sum_{m_{\Sigma_{1}}=1}^{l}\cdots\sum_{m_{\Sigma_{L-1}}=1}^{l}\sum_{m_{\Sigma_{L+1}}=1}^{l}\cdots\sum_{m_{\Sigma_{N}}=1}^{l}|\beta_{m_{\Sigma_{1}}\cdots m_{\Sigma_{L-1}}im_{\Sigma_{L+1}}\cdots m_{\Sigma_{N}}}|^{2}.
\end{eqnarray}
Clearly, these are the values that we wish to assign as the probabilities $p_{i}^{(N,l)}$ of measuring the outcomes corresponding to the states $|i\rangle$.

\subsubsection{Second Subcase}
The set \eqref{povmNN0l} of operators constitutes a POVM: each element $\mathcal{Q}_{i}^{(N,M,l)}$ meets the positivity condition \eqref{positivity} since
\begin{eqnarray}
\lefteqn{\langle\psi^{(N,l)}|\tilde{\mathcal{P}}_{i}^{\Sigma_{L}}|\psi^{(N,l)}\rangle=}\nonumber\\ &&\sum_{m_{\Sigma_{1}}=1}^{l}\cdots\sum_{m_{\Sigma_{L-1}}=1}^{l}\sum_{m_{\Sigma_{L+1}}=1}^{l}\cdots\sum_{m_{\Sigma_{N}}=1}^{l}|\beta_{m_{\Sigma_{1}}\cdots m_{\Sigma_{L-1}}im_{\Sigma_{L+1}}\cdots m_{\Sigma_{N}}}|^{2}\geq0
\end{eqnarray}
for all $L\leq M$, and together the elements meet the completeness condition \eqref{resolution} owing to \eqref{resolutionNN0l}. 

Now, with the probabilities $p_{i}^{(N,M,l)}$ given by the expectation values of $\mathcal{Q}_{i}^{(N,M,l)}$, I observe that this POVM respects the no extra vision principle: the expectation values of its elements $\mathcal{Q}_{i}^{(N,M,l)}$ are zero in the states \eqref{nevpstateNla}, the expectation values of its elements $\mathcal{Q}_{i}^{(N,M,l)}$ and $\mathcal{Q}_{j}^{(N,M,l)}$ are zero in the states \eqref{nevpstateNlb}, the expectation values of its elements $\mathcal{Q}_{i}^{(N,M,l)}$, $\mathcal{Q}_{j}^{(N,M,l)}$, and $\mathcal{Q}_{k}^{(N,M,l)}$ are zero in the states \eqref{nevpstateNlc}, \emph{et cetera}. As a convex combination of the projection operators $\tilde{\mathscr{P}}_{i}^{\Sigma_{L}}$ over the subset of $M$ identical copies, the POVM element $\mathcal{Q}_{i}^{(N,M,l)}$ possesses for each copy $L$ a term containing the projection operator $\mathcal{P}_{i}^{\Sigma_{L}}$. Thus, when acting on a state with the measurement outcome corresponding to the state $|i\rangle$ nowhere represented, the POVM element $\mathcal{Q}_{i}^{(N,M,l)}$ annihilates the state.

Next, I observe that the above POVM respects the probability symmetry principle: for each of the states \eqref{pspstateNla}, \eqref{pspstateNlb}, \eqref{pspstateNlc}, \emph{et cetera}, the expectation values of its elements $\mathcal{Q}_{i}^{(N,M,l)}$ and $\mathcal{Q}_{j}^{(N,M,l)}$ are equal. When acting on any of these states, both $\mathcal{Q}_{i}^{(N,M,l)}$ and $\mathcal{Q}_{j}^{(N,M,l)}$ select $n_{ij}$ times each and every component state $|m_{\Sigma_{1}}\cdots m_{\Sigma_{N}}\rangle$, yielding an expectation value of $\frac{n_{ij}}{M}$ for both POVM elements. 

Finally, consider the expectation value of $\mathcal{Q}_{i}^{(N,M,l)}$ in the state $|\psi^{(N,l)}\rangle$:
\begin{eqnarray}
\lefteqn{\langle\psi^{(N,l)}|\mathcal{Q}_{i}^{(N,M,l)}|\psi^{(N,l)}\rangle=}\nonumber\\ &&\frac{1}{M}\sum_{L=1}^{M}\sum_{m_{\Sigma_{1}}=1}^{l}\cdots\sum_{m_{\Sigma_{L-1}}=1}^{l}\sum_{m_{\Sigma_{L+1}}=1}^{l}\cdots\sum_{m_{\Sigma_{N}}=1}^{l}|\beta_{m_{\Sigma_{1}}\cdots m_{\Sigma_{L-1}}im_{\Sigma_{L+1}}\cdots m_{\Sigma_{N}}}|^{2}.
\end{eqnarray}
Clearly, these are the values that we wish to assign as the probabilities $p_{i}^{(N,M,l)}$ of measuring the outcomes corresponding to the states $|i\rangle$.

\section{On ``The Most General Case"}

\subsection{Formulating the No Extra Vision Principle}
I here make precise the meaning of the no extra vision principle for the setting of $\S 2.2.3$. I identify $l-1$ classes of states contained in the Hilbert space $\mathcal{H}^{(\{N\},l)}$ for which some of the potential measurement outcomes corresponding to the states $|i\rangle$ are not represented. I notate these classes of states, labeling them by the outcomes not present, as follows:
\begin{subequations}
\begin{eqnarray}
|\psi_{\neg i}^{(\{N\},l)}\rangle=\sum_{N=1}^{\mathscr{N}}\alpha_{N}\sum_{\substack{m_{\Sigma_{1}}=1\\ m_{\Sigma_{1}}\neq i}}^{l}\cdots\sum_{\substack{m_{\Sigma_{N}}=1\\ m_{\Sigma_{N}}\neq i}}^{l}\beta_{m_{\Sigma_{1}}\cdots m_{\Sigma_{N}}}|m_{\Sigma_{1}}\cdots m_{\Sigma_{N}}\rangle\label{nevpstatealephla}\\
|\psi_{\neg i \neg j}^{(\{N\},l)}\rangle=\sum_{N=1}^{\mathscr{N}}\alpha_{N}\sum_{\substack{m_{\Sigma_{1}}=1\\ m_{\Sigma_{1}}\neq i\\ m_{\Sigma_{1}}\neq j}}^{l}\cdots\sum_{\substack{m_{\Sigma_{N}}=1\\ m_{\Sigma_{N}}\neq i\\ m_{\Sigma_{N}}\neq j}}^{l}\beta_{m_{\Sigma_{1}}\cdots m_{\Sigma_{N}}}|m_{\Sigma_{1}}\cdots m_{\Sigma_{N}}\rangle\label{nevpstatealephlb}\\
|\psi_{\neg i \neg j \neg k}^{(\{N\},l)}\rangle=\sum_{N=1}^{\mathscr{N}}\alpha_{N}\sum_{\substack{m_{\Sigma_{1}}=1\\ m_{\Sigma_{1}}\neq i\\ m_{\Sigma_{1}}\neq j\\ m_{\Sigma_{1}}\neq k}}^{l}\cdots\sum_{\substack{m_{\Sigma_{N}}=1\\ m_{\Sigma_{N}}\neq i\\ m_{\Sigma_{N}}\neq j \\ m_{\Sigma_{N}}\neq k}}^{l}\beta_{m_{\Sigma_{1}}\cdots m_{\Sigma_{N}}}|m_{\Sigma_{1}}\cdots m_{\Sigma_{N}}\rangle\label{nevpstatealephlc}\\
et\, cetera\nonumber
\end{eqnarray}
\end{subequations}
for inequivalent $i,j,k$ \emph{et cetera} taking values in the set $\{1,\ldots,l\}$. Each successive class of states is contained within the previous class; for clarity in presenting the no extra vision principle's consequences, I find this classification most useful. This principle then dictates that the probabilities $p_{i}^{(\{N\},l)}$ of measuring the outcomes corresponding to the states $|i\rangle$ be zero in the states \eqref{nevpstatealephla}, that the probabilities $p_{i}^{(\{N\},l)}$ and $p_{j}^{(\{N\},l)}$ of respectively measuring the outcomes corresponding to the states $|i\rangle$ and $|j\rangle$ be zero in the states \eqref{nevpstatealephlb}, that the probabilities $p_{i}^{(\{N\},l)}$, $p_{j}^{(\{N\},l)}$, and $p_{k}^{(\{N\},l)}$ of respectively measuring the outcomes corresponding to the states $|i\rangle$, $|j\rangle$, and $|k\rangle$ be zero in the states \eqref{nevpstatealephlc}, \emph{et cetera}. I observe that I could have separately imposed the no extra vision principle within all branches of the universal state; this approach would have entailed the above formulation.

\subsection{Formulating the Probability Symmetry Principle}
I now make precise the meaning of the probability symmetry principle for the setting of $\S 2.2.3$. Since each branch of the universal state contains a different number of copies of the identical system, I cannot formulate this principle as straightforwardly as above. Taking a hint from the closing observation of $\S\mathrm{B}.1$, I determine that we must separately impose the probability symmetry principle within each branch of the universal state. I identify $\prod_{N=2}^{\mathscr{N}}\lfloor\frac{N}{2}\rfloor$ classes of states contained in the Hilbert space $\mathcal{H}^{(\{N\},l)}$ for which within each branch the potential measurement outcomes corresponding to the states $|i\rangle$ and $|j\rangle$ are represented in equal numbers.\footnote{The probability symmetry principle only applies to branches for which $N\geq2$.} I notate these states, labeling them by the numbers $n_{ij}^{N}$ of occurrences of each outcome within each branch, as follows:
\begin{subequations}
\begin{eqnarray}
\lefteqn{|\psi_{n_{ij}^{2}=1,n_{ij}^{3}=1,\ldots,n_{ij}^{\mathscr{N}}=1}^{(\{N\},l)}\rangle=}\nonumber\\ &&\alpha_{2}|\psi_{n_{ij}=1}^{(2,l)}\rangle+\alpha_{3}|\psi_{n_{ij}=1}^{(3,l)}\rangle+\alpha_{4}|\psi_{n_{ij}=1}^{(4,l)}\rangle+\cdots+\alpha_{\mathscr{N}}|\psi_{n_{ij}=1}^{(\mathscr{N},l)}\rangle\label{pspstatealephla}\\
\lefteqn{|\psi_{n_{ij}^{2}=1,n_{ij}^{3}=2,n_{ij}^{4}=1,\ldots,n_{ij}^{\mathscr{N}}=1}^{(\{N\},l)}\rangle=}\nonumber\\ &&\alpha_{2}|\psi_{n_{ij}=1}^{(2,l)}\rangle+\alpha_{3}|\psi_{n_{ij}=1}^{(3,l)}\rangle+\alpha_{4}|\psi_{n_{ij}=2}^{(4,l)}\rangle+\alpha_{5}|\psi_{n_{ij}=1}^{(5,l)}\rangle+\cdots+\alpha_{\mathscr{N}}|\psi_{n_{ij}=1}^{(\mathscr{N},l)}\rangle\label{pspstatealephlb}\\
\lefteqn{|\psi_{n_{ij}^{2}=1,n_{ij}^{3}=1,n_{ij}^{4}=2,n_{ij}^{5}=1,\ldots,n_{ij}^{\mathscr{N}}=1}^{(\{N\},l)}\rangle=}\nonumber\\ &&\alpha_{2}|\psi_{n_{ij}=1}^{(2,l)}\rangle+\alpha_{3}|\psi_{n_{ij}=1}^{(3,l)}\rangle+\alpha_{4}|\psi_{n_{ij}=1}^{(4,l)}\rangle+\alpha_{5}|\psi_{n_{ij}=2}^{(5,l)}\rangle+\alpha_{6}|\psi_{n_{ij}=1}^{(6,l)}\rangle+\cdots+\alpha_{\mathscr{N}}|\psi_{n_{ij}=1}^{(\mathscr{N},l)}\rangle\label{pspstatealephlc}\\
et\, cetera\nonumber
\end{eqnarray}
\end{subequations}
for inequivalent $i,j\in\{1,\ldots,l\}$.\footnote{I do not allow $n_{ij}^{N}=0$ since, for this class of states within a given branch, the probability symmetry principle is equivalent to the no extra vision principle.} The states $|\psi_{n_{ij}}^{(N,l)}\rangle$ are those listed in \eqref{pspstateNla}, \eqref{pspstateNlb}, \eqref{pspstateNlc}, \emph{et cetera}. Each successive class of states contains the previous class; for clarity in presenting the probability symmetry principle's consequences, I find this classification most useful. This principle now dictates that, for each state within the classes above, the probabilities $p_{i}^{(\{N\},l)}$ and $p_{j}^{(\{N\},l)}$ of respectively measuring the outcomes corresponding to the states $|i\rangle$ and $|j\rangle$ be equal.

\subsection{Properties of the Constructed Positive Operator Valued Measures}

\subsubsection{First Subcase}
The set \eqref{povmalephl} of operators constitutes a POVM: each elements $\mathcal{Q}_{i}^{(\{N\},l)}$ meets the positivity condition \eqref{positivity} since, for any particular $N$, 
\begin{eqnarray}
\lefteqn{\langle\psi^{(\{N\},l)}|\mathscr{P}_{i}^{\Sigma_{L}}|\psi^{(\{N\},l)}\rangle=}\nonumber\\ &&|\alpha_{N}|^{2}\sum_{m_{\Sigma_{1}}=1}^{l}\cdots\sum_{m_{\Sigma_{L-1}}=1}^{l}\sum_{m_{\Sigma_{L+1}}=1}^{l}\cdots\sum_{m_{\Sigma_{N}}=1}^{l}|\beta_{m_{\Sigma_{1}}\cdots m_{\Sigma_{L-1}}im_{\Sigma_{L+1}}\cdots m_{\Sigma_{N}}}|^{2}\geq 0
\end{eqnarray}
for all $L$, and together the elements meet the completeness condition \eqref{resolution} owing to \eqref{resolutionalephl}.

Now, with the probabilities $p_{i}^{(\{N\},l)}$ given by the expectation values of $\mathcal{Q}_{i}^{(\{N\},l)}$, I observe that this POVM respects the no extra vision principle: the expectation values of its elements $\mathcal{Q}_{i}^{(\{N\},l)}$ are zero in the states \eqref{nevpstatealephla}, the expectation values of its elements $\mathcal{Q}_{i}^{(\{N\},l)}$ and $\mathcal{Q}_{j}^{(\{N\},l)}$ are zero in the states \eqref{nevpstatealephlb}, the expectation values of its elements $\mathcal{Q}_{i}^{(\{N\},l)}$, $\mathcal{Q}_{j}^{(\{N\},l)}$, and $\mathcal{Q}_{k}^{(\{N\},l)}$ are zero in the states \eqref{nevpstatealephlc}, \emph{et cetera}. As a convex combination of the projection operators $\mathscr{P}_{i}^{\Sigma_{L}}$ over all N identical copies of each branch, the POVM element $\mathcal{Q}_{i}^{(\{N\},l)}$ possesses for each copy $L$ of each branch a term containing the projection operator $\mathcal{P}_{i}^{\Sigma_{L}}$. Thus, when acting on a state with the measurement outcome corresponding to the state $|i\rangle$ nowhere represented, the POVM element $\mathcal{Q}_{i}^{(\{N\},l)}$ annihilates the state.

Next, I observe that this POVM respects the probability symmetry principle: for each of the states \eqref{pspstatealephla}, \eqref{pspstatealephlb}, \eqref{pspstatealephlc}, \emph{et cetera}, the expectation values of its elements $\mathcal{Q}_{i}^{(\{N\},l)}$ and $\mathcal{Q}_{j}^{(\{N\},l)}$ are equal. When acting on any of these states, within each branch both $\mathcal{Q}_{i}^{(\{N\},l)}$ and $\mathcal{Q}_{j}^{(\{N\},l)}$ select $n_{ij}^{N}$ times each and every component state $|m_{\Sigma_{1}}\cdots m_{\Sigma_{N}}\rangle$, yielding an expectation value of $\sum_{N=2}^{\mathscr{N}}|\alpha_{N}|^{2}\frac{n_{ij}^{N}}{N}$ for both POVM elements.

Finally, consider the expectation value of $\mathcal{Q}_{i}^{(\{N\},l)}$ in the state $|\psi^{(\{N\},l)}\rangle$:
\begin{eqnarray}
\lefteqn{\langle\psi^{(\{N\},l)}|\mathcal{Q}_{i}^{(\{N\},l)}|\psi^{(\{N\},l)}\rangle=}\nonumber\\ &&\sum_{N=1}^{\mathscr{N}}|\alpha_{N}|^{2}\frac{1}{N}\sum_{L=1}^{N}\sum_{m_{\Sigma_{1}}=1}^{l}\cdots\sum_{m_{\Sigma_{L-1}}=1}^{l}\sum_{m_{\Sigma_{L+1}}=1}^{l}\cdots\sum_{m_{\Sigma_{N}}=1}^{l}|\beta_{m_{\Sigma_{1}}\cdots m_{\Sigma_{L-1}}im_{\Sigma_{L+1}}\cdots m_{\Sigma_{N}}}|^{2}.
\end{eqnarray}
Clearly, these are the values that we wish to assign as the probabilities $p_{i}^{(\{N\},l)}$ of measuring the outcomes corresponding to the states $|i\rangle$.

\subsubsection{Second Subcase}
The set \eqref{povmaleph0l} of operators constitutes a POVM: each elements $\mathcal{Q}_{i}^{(\{N\},\{M\},l)}$ meets the positivity condition \eqref{positivity} since, for any particular $N$,
\begin{eqnarray}
\lefteqn{\langle\psi^{(\{N\},l)}|\tilde{\mathscr{P}}_{i}^{\Sigma_{L}}|\psi^{(\{N\},l)}\rangle=}\nonumber\\ &&|\alpha_{N}|^{2}\sum_{m_{\Sigma_{1}}=1}^{l}\cdots\sum_{m_{\Sigma_{L-1}}=1}^{l}\sum_{m_{\Sigma_{L+1}}=1}^{l}\cdots\sum_{m_{\Sigma_{N}}=1}^{l}|\beta_{m_{\Sigma_{1}}\cdots m_{\Sigma_{L-1}}im_{\Sigma_{L+1}}\cdots m_{\Sigma_{N}}}|^{2}\geq 0
\end{eqnarray}
for all $L<M_{N}$, and together the elements meet the completeness condition \eqref{resolution} owing to \eqref{resolutionaleph0l}.

Now, with the probabilities $p_{i}^{(\{N\},\{M\},l)}$ given by the expectation values of $\mathcal{Q}_{i}^{(\{N\},\{M\},l)}$, I observe that this POVM respects the no extra vision principle: the expectation values of its elements $\mathcal{Q}_{i}^{(\{N\},\{M\},l)}$ are zero in the states \eqref{nevpstatealephla}, the expectation values of its elements $\mathcal{Q}_{i}^{(\{N\},\{M\},l)}$ and $\mathcal{Q}_{j}^{(\{N\},\{M\},l)}$ are zero in the states \eqref{nevpstatealephlb}, the expectation values of its elements $\mathcal{Q}_{i}^{(\{N\},\{M\},l)}$, $\mathcal{Q}_{j}^{(\{N\},\{M\},l)}$, and $\mathcal{Q}_{k}^{(\{N\},\{M\},l)}$ are zero in the states \eqref{nevpstatealephlc}, \emph{et cetera}. As a convex combination of the projection operators $\mathscr{P}_{i}^{\Sigma_{L}}$ over the subsets of $M_{N}$ identical copies of each branch, the POVM element $\mathcal{Q}_{i}^{(\{N\},\{M\},l)}$ possesses for each copy $L$ of each branch a term containing the projection operator $\mathcal{P}_{i}^{\Sigma_{L}}$. Thus, when acting on a state with the measurement outcome corresponding to the state $|i\rangle$ nowhere represented, the POVM element $\mathcal{Q}_{i}^{(\{N\},\{M\},l)}$ annihilates the state.

Next, I observe that this POVM respects the probability symmetry principle:  for each of the states \eqref{pspstatealephla}, \eqref{pspstatealephlb}, \eqref{pspstatealephlc}, \emph{et cetera}, the expectation values of its elements $\mathcal{Q}_{i}^{(\{N\},\{M\},l)}$ and $\mathcal{Q}_{j}^{(\{N\},\{M\},l)}$ are equal. When acting on any of these states, within each branch both $\mathcal{Q}_{i}^{(\{N\},\{M\},l)}$ and $\mathcal{Q}_{j}^{(\{N\},\{M\},l)}$ select $n_{ij}^{N}$ times each and every component state $|m_{\Sigma_{1}}\cdots m_{\Sigma_{N}}\rangle$, yielding an expectation value of $\sum_{N=2}^{\mathscr{N}}|\alpha_{N}|^{2}\frac{n_{ij}^{N}}{M_{N}}$ for both POVM elements.
 
Finally, consider the expectation value of $\mathcal{Q}_{i}^{(\{N\},\{M\},l)}$ in the state $|\psi^{(\{N\},l)}\rangle$:
\begin{eqnarray}
\lefteqn{\langle\psi^{(\{N\},l)}|\mathcal{Q}_{i}^{(\{N\},\{M\},l)}|\psi^{(\{N\},l)}\rangle=}\nonumber\\ &&\sum_{N=1}^{\mathscr{N}}|\alpha_{N}|^{2}\frac{1}{M_{N}}\sum_{L=1}^{M_{N}}\sum_{m_{\Sigma_{1}}=1}^{l}\cdots\sum_{m_{\Sigma_{L-1}}=1}^{l}\sum_{m_{\Sigma_{L+1}}=1}^{l}\cdots\sum_{m_{\Sigma_{N}}=1}^{l}|\beta_{m_{\Sigma_{1}}\cdots m_{\Sigma_{L-1}}im_{\Sigma_{L+1}}\cdots m_{\Sigma_{N}}}|^{2}.
\end{eqnarray}
Clearly, these are the values that we wish to assign as the probabilities $p_{i}^{(\{N\},\{M\},l)}$ of measuring the outcomes corresponding to the states $|i\rangle$.

\section{On Permutation Symmetries in a Very Small Universe}

Consider again the universe of $\S 1$ containing two copies of an exactly identical system at different spacetime locations each of which has associated to it a $2$-dimensional Hilbert space. \footnote{I revert to the notation of $\S 1$.} The $4$-dimensional Hilbert space $\mathcal{H}=\mathcal{H}_{\Sigma}\otimes\mathcal{H}_{\Omega}$ of these two copies decomposes as $\mathcal{H}=\mathcal{H}_{\mathcal{A}}\oplus\mathcal{H}_{\mathcal{S}}$ under the action of the symmetric group $S_{2}$. Here, $\mathcal{H}_{\mathcal{A}}$ is the Hilbert subspace spanned by the singlet state
\begin{equation}
|\psi_{12}\rangle_{\mathcal{A}}=\frac{1}{\sqrt{2}}\left(|12\rangle-|21\rangle\right),\label{singlet}
\end{equation}
which is invariant under the antisymmetric representation of $S_{2}$, and $\mathcal{H}_{\mathcal{S}}$ is the Hilbert subspace spanned by the triplet states
\begin{equation}
|\psi_{11}\rangle_{\mathcal{S}}=|11\rangle,\quad|\psi_{12}\rangle_{\mathcal{S}}=\frac{1}{\sqrt{2}}\left(|12\rangle+|21\rangle\right),\quad|\psi_{22}\rangle_{\mathcal{S}}=|22\rangle,\label{triplet}
\end{equation}
which are invariant under the symmetric representation of $S_{2}$. 

\subsection{Imposing the Superselection Rule for Symmetry Type}

\subsubsection{On the Nonexistence of a Complete Set of Orthogonal Projection Operators}

I now attempt to construct a complete set $\{\mathcal{P}_{1},\mathcal{P}_{2}\}$ of orthogonal projection operators respecting the superselection rule for symmetry type the expectation values of whose elements yield the probabilities $p_{1}$ and $p_{2}$ of respectively measuring the outcomes corresponding to the states $|1\rangle$ and $|2\rangle$. The superselection rule dictates that any operator corresponding to a physical observable be block diagonal with respect to the decomposition of the Hilbert space $\mathcal{H}$ into the superselection sectors $\mathcal{H}_{\mathcal{A}}$ and $\mathcal{H}_{\mathcal{S}}$. In other words, these operators must have vanishing matrix elements between states from the two superselection sectors. Employing the singlet-triplet basis, the desired projection operators thus assume the following forms:
\begin{eqnarray}
\mathcal{P}_{1}&=&\omega_{11}|\psi_{12}\rangle_{\mathcal{A}\mathcal{A}}\langle\psi_{12}|\nonumber\\ &&+ \omega_{22}|\psi_{11}\rangle_{\mathcal{S}\mathcal{S}}\langle\psi_{11}|+ \omega_{23}|\psi_{11}\rangle_{\mathcal{S}\mathcal{S}}\langle\psi_{12}|+ \omega_{24}|\psi_{11}\rangle_{\mathcal{S}\mathcal{S}}\langle\psi_{22}|\nonumber\\ &&+ \omega_{32}|\psi_{12}\rangle_{\mathcal{S}\mathcal{S}}\langle\psi_{11}|+ \omega_{33}|\psi_{12}\rangle_{\mathcal{S}\mathcal{S}}\langle\psi_{12}|+ \omega_{34}|\psi_{12}\rangle_{\mathcal{S}\mathcal{S}}\langle\psi_{22}|\nonumber\\ &&+ \omega_{42}|\psi_{22}\rangle_{\mathcal{S}\mathcal{S}}\langle\psi_{11}|+ \omega_{43}|\psi_{22}\rangle_{\mathcal{S}\mathcal{S}}\langle\psi_{12}|+ \omega_{44}|\psi_{22}\rangle_{\mathcal{S}\mathcal{S}}\langle\psi_{22}|
\end{eqnarray}
for complex coefficients $\omega_{ij}$ and
\begin{eqnarray}
\mathcal{P}_{2}&=&\chi_{11}|\psi_{12}\rangle_{\mathcal{A}\mathcal{A}}\langle\psi_{12}|\nonumber\\ &&+ \chi_{22}|\psi_{11}\rangle_{\mathcal{S}\mathcal{S}}\langle\psi_{11}|+ \chi_{23}|\psi_{11}\rangle_{\mathcal{S}\mathcal{S}}\langle\psi_{12}|+ \chi_{24}|\psi_{11}\rangle_{\mathcal{S}\mathcal{S}}\langle\psi_{22}|\nonumber\\ &&+ \chi_{32}|\psi_{12}\rangle_{\mathcal{S}\mathcal{S}}\langle\psi_{11}|+ \chi_{33}|\psi_{12}\rangle_{\mathcal{S}\mathcal{S}}\langle\psi_{12}|+ \chi_{34}|\psi_{12}\rangle_{\mathcal{S}\mathcal{S}}\langle\psi_{22}|\nonumber\\ &&+ \chi_{42}|\psi_{22}\rangle_{\mathcal{S}\mathcal{S}}\langle\psi_{11}|+ \chi_{43}|\psi_{22}\rangle_{\mathcal{S}\mathcal{S}}\langle\psi_{12}|+ \chi_{44}|\psi_{22}\rangle_{\mathcal{S}\mathcal{S}}\langle\psi_{22}|
\end{eqnarray}
for complex coefficients $\chi_{ij}$. Owing to the required Hermiticity of these projection operators, the coefficients satisfy the following relations:
\begin{equation}
\omega_{11}=w_{11},\quad \omega_{22}=w_{22},\quad \omega_{33}=w_{33},\quad \omega_{44}=w_{44}
\end{equation}
for real parameters $w_{ii}$,
\begin{equation}
\omega_{23}= \omega_{32}^{*},\quad \omega_{24}= \omega_{42}^{*},\quad \omega_{34}= \omega_{43}^{*},
\end{equation}
and
\begin{equation}
\chi_{11}=x_{11},\quad \chi_{22}=x_{22},\quad \chi_{33}=x_{33},\quad \chi_{44}=x_{44}
\end{equation}
for real parameters $x_{ii}$,
\begin{equation}
\chi_{23}= \chi_{32}^{*},\quad \chi_{24}= \chi_{42}^{*},\quad \chi_{34}= \chi_{43}^{*}.
\end{equation}

Now, I insist that the pair $\{\mathcal{P}_{1},\mathcal{P}_{2}\}$ of operators satisfy the conditions of idempotency, orthogonality, and completeness expressed in \eqref{completeness} and \eqref{idemortho}. Idempotency imposes ten real relations on each of the sets $\{\omega_{ij}\}$ and $\{\chi_{ij}\}$ of complex parameters respectively characterizing the operators $\mathcal{P}_{1}$ and $\mathcal{P}_{2}$. In particular, I find that 
\begin{subequations}
\begin{eqnarray}
w_{11}^{2}=w_{11}\\ x_{11}^{2}=x_{11}
\end{eqnarray}
\end{subequations}
relations only satisfied for $w_{11}=1$ and $x_{11}=1$. These requirements immediately contradict the condition of completeness: the coefficient of $|\psi_{12}\rangle_{\mathcal{A}\mathcal{A}}\langle\psi_{12}|$ in the sum $\mathcal{P}_{1}+\mathcal{P}_{2}$, given here by $w_{11}+x_{11}$, must be unity. Evidently, there does not exist a pair of projection operators respecting the superselection rule for symmetry type satisfying the desired criteria.

\subsubsection{On the Existence of a Positive Operator Valued Measure}

There does exist a POVM respecting the superselection rule for symmetry type the expectation values of whose elements yield the probabilities $p_{1}$ and $p_{2}$ of respectively measuring the outcomes corresponding to the states $|1\rangle$ and $|2\rangle$. The POVM of $\S 2.2.1$ given in \eqref{povm22a} and \eqref{povm22b} has the following form in the singlet-triplet basis:
\begin{subequations}
\begin{eqnarray}
\mathcal{Q}_{1}=\frac{1}{2}|\psi_{12}\rangle_{\mathcal{A}\mathcal{A}}\langle\psi_{12}|+|\psi_{11}\rangle_{\mathcal{S}\mathcal{S}}\langle\psi_{11}|+\frac{1}{2}|\psi_{12}\rangle_{\mathcal{S}\mathcal{S}}\langle\psi_{12}|\\
\mathcal{Q}_{2}=\frac{1}{2}|\psi_{12}\rangle_{\mathcal{A}\mathcal{A}}\langle\psi_{12}|+\frac{1}{2}|\psi_{12}\rangle_{\mathcal{S}\mathcal{S}}\langle\psi_{12}|+|\psi_{22}\rangle_{\mathcal{S}\mathcal{S}}\langle\psi_{22}|
\end{eqnarray}
\end{subequations}
Clearly, this POVM is block diagonal with respect to the decomposition of the Hilbert space $\mathcal{H}$ into the superselection sectors $\mathcal{H}_{\mathcal{A}}$ and $\mathcal{H}_{\mathcal{S}}$. Moreover, as demonstrated in $\S 2.2.1$, this POVM satisfies all of the desired criteria.

\subsection{Imposing the Symmetrization Postulate}

\subsubsection{On the Nonexistence of a Complete Set of Orthogonal Projection Operators}

The contradiction that I derived in $\S\mathrm{C}.1.1$ arose within the Hilbert subspace $\mathcal{H}_{\mathcal{A}}$, thereby preventing the construction of the desired projection operators on the full Hilbert space $\mathcal{H}$. Imposing the symmetrization postulate, which instructs us to take either $\mathcal{H}_{\mathcal{A}}$ or $\mathcal{H}_{\mathcal{S}}$ as the fundamental Hilbert space, the circumvention of this contradiction seems at least a possibility. To investigate this possibility, I attempt to construct a complete set $\{\mathcal{P}_{1},\mathcal{P}_{2}\}$ of orthogonal projection operators respecting the symmetrization postulate the expectation values of whose elements yield the probabilities $p_{1}$ and $p_{2}$ of respectively measuring the outcomes corresponding to the states $|1\rangle$ and $|2\rangle$. I analyze in turn the two cases---$\mathcal{H}_{\mathcal{A}}$ as the fundamental Hilbert space and $\mathcal{H}_{\mathcal{S}}$ as the fundamental Hilbert space.

With $\mathcal{H}_{\mathcal{A}}$ as the Hilbert space, there is no possibility of constructing the set $\{\mathcal{P}_{1},\mathcal{P}_{2}\}$ of projection operators. Since this Hilbert space is $1$-dimensional, there exists only one projection operator, namely $|\psi_{12}\rangle_{\mathcal{A}\mathcal{A}}\langle\psi_{12}|$, which of course yields unity for its expectation value in the state $|\psi_{12}\rangle_{\mathcal{A}}$ spanning $\mathcal{H}_{\mathcal{A}}$.

With $\mathcal{H}_{\mathcal{S}}$ as the Hilbert space, the desired projection operators assume the following forms in the triplet basis:
\begin{eqnarray}
\mathcal{P}_{1}&=&\upsilon_{11}|\psi_{11}\rangle_{\mathcal{S}\mathcal{S}}\langle\psi_{11}|+ \upsilon_{12}|\psi_{11}\rangle_{\mathcal{S}\mathcal{S}}\langle\psi_{12}|+ \upsilon_{13}|\psi_{11}\rangle_{\mathcal{S}\mathcal{S}}\langle\psi_{22}|\nonumber\\ &&+ \upsilon_{21}|\psi_{12}\rangle_{\mathcal{S}\mathcal{S}}\langle\psi_{11}|+ \upsilon_{22}|\psi_{12}\rangle_{\mathcal{S}\mathcal{S}}\langle\psi_{12}|+ \upsilon_{23}|\psi_{12}\rangle_{\mathcal{S}\mathcal{S}}\langle\psi_{22}|\nonumber\\ &&+ \upsilon_{31}|\psi_{22}\rangle_{\mathcal{S}\mathcal{S}}\langle\psi_{11}|+ \upsilon_{32}|\psi_{22}\rangle_{\mathcal{S}\mathcal{S}}\langle\psi_{12}|+ \upsilon_{33}|\psi_{22}\rangle_{\mathcal{S}\mathcal{S}}\langle\psi_{22}|
\end{eqnarray}
for complex coefficients $\upsilon_{ij}$ and
\begin{eqnarray}
\mathcal{P}_{2}&=&\nu_{11}|\psi_{11}\rangle_{\mathcal{S}\mathcal{S}}\langle\psi_{11}|+ \nu_{12}|\psi_{11}\rangle_{\mathcal{S}\mathcal{S}}\langle\psi_{12}|+ \nu_{13}|\psi_{11}\rangle_{\mathcal{S}\mathcal{S}}\langle\psi_{22}|\nonumber\\ &&+ \nu_{21}|\psi_{12}\rangle_{\mathcal{S}\mathcal{S}}\langle\psi_{11}|+ \nu_{22}|\psi_{12}\rangle_{\mathcal{S}\mathcal{S}}\langle\psi_{12}|+ \nu_{23}|\psi_{12}\rangle_{\mathcal{S}\mathcal{S}}\langle\psi_{22}|\nonumber\\ &&+ \nu_{31}|\psi_{22}\rangle_{\mathcal{S}\mathcal{S}}\langle\psi_{11}|+ \nu_{32}|\psi_{22}\rangle_{\mathcal{S}\mathcal{S}}\langle\psi_{12}|+ \nu_{33}|\psi_{22}\rangle_{\mathcal{S}\mathcal{S}}\langle\psi_{22}|
\end{eqnarray}
for complex coefficients $\nu_{ij}$. Owing to the required Hermiticity of these projection operators, the coefficients satisfy the following relations:
\begin{equation}
\upsilon_{11}=u_{11},\quad \upsilon_{22}=u_{22},\quad \upsilon_{33}=u_{33}
\end{equation}
for real parameters $u_{ii}$,
\begin{equation}
\upsilon_{12}= \upsilon_{21}^{*},\quad \upsilon_{13}= \upsilon_{31}^{*},\quad \upsilon_{23}= \upsilon_{32}^{*},
\end{equation}
and
\begin{equation}
\nu_{11}=v_{11},\quad \nu_{22}=v_{22},\quad \nu_{33}=v_{33}
\end{equation}
for real parameters $v_{ii}$,
\begin{equation}
\nu_{12}= \nu_{21}^{*},\quad \nu_{13}= \nu_{31}^{*},\quad \nu_{23}= \nu_{32}^{*}
\end{equation}

Now, I insist that the pair $\{\mathcal{P}_{1},\mathcal{P}_{2}\}$ of operators satisfy the conditions of idempotency, orthogonality, and completeness expressed in \eqref{completeness} and \eqref{idemortho}. Idempotency requires that
\begin{subequations}
\begin{eqnarray}
u_{11}=u_{11}^{2}+|\upsilon_{12}|^{2}+|\upsilon_{13}|^{2}\\ u_{22}=|\upsilon_{12}|^{2}+u_{22}^{2}+|\upsilon_{23}|^{2}\\ u_{33}=|\upsilon_{13}|^{2}+|\upsilon_{23}|^{2}+u_{33}^{2}\label{idem22spa}
\end{eqnarray}
\end{subequations}
and that
\begin{subequations}
\begin{eqnarray}
v_{11}=v_{11}^{2}+|\nu_{12}|^{2}+| \nu_{13}|^{2}\\ v_{22}=|\nu_{12}|^{2}+v_{22}^{2}+|\nu_{23}|^{2}\\v_{33}=| \nu_{13}|^{2}+| \nu_{23}|^{2}+v_{33}^{2}\label{idem22spb}
\end{eqnarray}
\end{subequations}
Orthogonality requires that
\begin{equation}
\upsilon_{12}^{*}\nu_{12}+u_{22}v_{22}+\upsilon_{23}\nu_{23}^{*}=0\label{ortho22sp}
\end{equation}
Completeness requires that
\begin{equation}
u_{22}+v_{22}=1\label{completeness22sp}
\end{equation}
The conditions \eqref{idem22spa}, \eqref{idem22spb}, \eqref{ortho22sp}, and \eqref{completeness22sp} represent only a subset of those entailed by idempotency, orthogonality, and completeness.

Next, since the projection operators $\mathcal{P}_{1}$ and $\mathcal{P}_{2}$ should yield the probabilities $p_{1}$ and $p_{2}$ as their respective expectation values, I insist at a minimum that
\begin{subequations}
\begin{eqnarray}
_{\mathcal{S}}\langle\psi_{11}|\mathcal{P}_{1}|\psi_{11}\rangle_{\mathcal{S}}=1,\quad _{\mathcal{S}}\langle\psi_{22}|\mathcal{P}_{1}|\psi_{22}\rangle_{\mathcal{S}}=0 \\ _{\mathcal{S}}\langle\psi_{11}|\mathcal{P}_{2}|\psi_{11}\rangle_{\mathcal{S}}=0,\quad _{\mathcal{S}}\langle\psi_{22}|\mathcal{P}_{2}|\psi_{22}\rangle_{\mathcal{S}}=1
\end{eqnarray}
\end{subequations}
These conditions dictate that
\begin{subequations}
\begin{eqnarray}
u_{11}=1,\quad u_{33}=0\\
v_{11}=0,\quad v_{33}=1
\end{eqnarray}
\end{subequations}
Combined with the conditions expressed in \eqref{idem22spa} and \eqref{idem22spb}, I conclude first that
\begin{subequations}
\begin{eqnarray}
\upsilon_{12}=0,\quad \upsilon_{13}=0, \quad \upsilon_{23}=0\\
\nu_{12}=0,\quad \nu_{13}=0, \quad \nu_{23}=0
\end{eqnarray}
\end{subequations}
and then that
\begin{subequations}
\begin{eqnarray}
u_{22}=1\\ v_{22}=1
\end{eqnarray}
\end{subequations}
This last consequence represents an immediate contradiction of \eqref{completeness22sp}. Evidently, there does not exist a pair of projection operators respecting the symmetrization postulate satisfying the desired criteria.

\subsubsection{On the Existence of a Positive Operator Valued Measure}

There does exist a POVM respecting the symmetrization postulate the expectation values of whose elements yield the probabilities $p_{1}$ and $p_{2}$ of respectively measuring the outcomes corresponding to the states $|1\rangle$ and $|2\rangle$. With $\mathcal{H}_{\mathcal{A}}$ as the Hilbert space, I take as the POVM elements
\begin{subequations}
\begin{eqnarray}
\mathcal{Q}_{1}=\frac{1}{2}|\psi_{12}\rangle_{\mathcal{A}\mathcal{A}}\langle\psi_{12}|\\
\mathcal{Q}_{2}=\frac{1}{2}|\psi_{12}\rangle_{\mathcal{A}\mathcal{A}}\langle\psi_{12}|
\end{eqnarray}
\end{subequations}
With $\mathcal{H}_{\mathcal{S}}$ as the Hilbert space, I take as the POVM elements
\begin{subequations}
\begin{eqnarray}
\mathcal{Q}_{1}=|\psi_{11}\rangle_{\mathcal{S}\mathcal{S}}\langle\psi_{11}|+\frac{1}{2}|\psi_{12}\rangle_{\mathcal{S}\mathcal{S}}\langle\psi_{12}|\\
\mathcal{Q}_{2}=\frac{1}{2}|\psi_{12}\rangle_{\mathcal{S}\mathcal{S}}\langle\psi_{12}|+|\psi_{22}\rangle_{\mathcal{S}\mathcal{S}}\langle\psi_{22}|
\end{eqnarray}
\end{subequations}
Clearly, these two POVMs satisfy all of the desired criteria.

\end{document}